\documentclass[12pt]{article}
\usepackage[dvips]{graphicx}
\usepackage{xspace}
\usepackage{epsfig}
\usepackage{amssymb}
\usepackage{cite}
\usepackage{./axodraw}
 \newcounter{enumct}
 \newenvironment{Enumerate}{\begin{list}{\arabic{enumct}.}%
 {\usecounter{enumct}\setlength{\topsep}{0.2mm}%
 \setlength{\partopsep}{0.2mm}\setlength{\itemsep}{0.2mm}%
 \setlength{\parsep}{0.2mm}}}{\end{list}}
\newcommand{\aem    }{\ensuremath{\alpha}\xspace}
\newcommand{\aemsq  }{\ensuremath{\aem^2}\xspace}
\newcommand{\invpb  }{\ensuremath{\rm pb^{-1}}\xspace}
\newcommand{\msq    }{\ensuremath{m_\mu^{2}}\xspace}
\newcommand{\wsq    }{\ensuremath{W^{2}}\xspace}
\newcommand{\qsq    }{\ensuremath{Q^{2}}\xspace}
\newcommand{\psq    }{\ensuremath{P^{2}}\xspace}
\newcommand{\sqsq   }{\ensuremath{q^{2}}\xspace}
\newcommand{\spsq   }{\ensuremath{p^{2}}\xspace}
\newcommand{\psqmin }{\ensuremath{P^{2}_{\rm{min}}}\xspace}
\newcommand{\psqmax }{\ensuremath{P^{2}_{\rm{max}}}\xspace}
\newcommand{\gev    }{\ensuremath{\rm GeV}\xspace}
\newcommand{\gevsq  }{\ensuremath{\rm GeV^2}\xspace}
\newcommand{\zn     }{\ensuremath{\rm Z}\xspace}
\newcommand{\epem   }{\ensuremath{\rm e^+e^-}\xspace}
\newcommand{\ee     }{\ensuremath{\rm ee}\xspace}
\newcommand{\mupmum }{\ensuremath{\mu^+\mu^-}\xspace}
\newcommand{\ft     }{\ensuremath{F_{\mathrm{2,{\mathrm had}}}^{\gamma}}\xspace}
\newcommand{\ftqed  }{\ensuremath{F_{\mathrm{2}}^{\gamma}}\xspace}

\newcommand{\foqed  }{\ensuremath{F_{1}^{\gamma}}\xspace}
\newcommand{\flqed  }{\ensuremath{F_{\mathrm{L}}^{\gamma}}\xspace}
\newcommand{\faqed  }{\ensuremath{F_{\mathrm{A}}^{\gamma}}\xspace}
\newcommand{\fbqed  }{\ensuremath{F_{\mathrm{B}}^{\gamma}}\xspace}
\newcommand{\ftxqp  }{\ensuremath{\ftqed(x,\qsq,\psq)}\xspace}
\newcommand{\flxqp  }{\ensuremath{\flqed(x,\qsq,\psq)}\xspace}
\newcommand{\zntau  }{\ensuremath{\zn\rightarrow \tau^+\tau^-}\xspace}
\newcommand{\znmu   }{\ensuremath{\zn\rightarrow \mu^+\mu^-}\xspace}
\newcommand{\zne    }{\ensuremath{\zn\rightarrow \epem}\xspace}
\newcommand{\gmbg   }{\ensuremath{\gamma^{\star}\gamma^{(\star)}}\xspace}

\newcommand{\gsg    }{\ensuremath{\gamma^{\star}\gamma}\xspace}
\newcommand{\ggss   }{\ensuremath{\gamma^{\star}\gamma^{\star}}\xspace}
\newcommand{\gghad  }{\ensuremath{\gamma\gamma^{\star}\rightarrow\mbox{hadrons}}\xspace}
\newcommand{\ggtau  }{\ensuremath{\gamma\gamma^{\star}\rightarrow\tau^+\tau^-}\xspace}
\newcommand{\ph     }{\ensuremath{\phi}\xspace}
\newcommand{\etag   }{\ensuremath{E_{\rm tag}}\xspace}
\newcommand{\estag  }{\ensuremath{E_{\rm stag}}\xspace}
\newcommand{\eone   }{\ensuremath{E_1}\xspace}
\newcommand{\etwo   }{\ensuremath{E_2}\xspace}

\newcommand{\eb     }{\ensuremath{E_{\rm b}}\xspace}
\newcommand{\emin   }{\ensuremath{{\rm min}(\eone,\etwo)}\xspace}
\newcommand{\az     }{\ensuremath{\chi}\xspace}
\newcommand{\thet   }{\ensuremath{\theta}\xspace}
\newcommand{\tmin   }{\ensuremath{{\rm min}(\tone,\ttwo)}\xspace}
\newcommand{\ttag   }{\ensuremath{\theta_{\rm tag}}\xspace}
\newcommand{\tstag  }{\ensuremath{\theta_{\rm stag}}\xspace}
\newcommand{\tone   }{\ensuremath{\theta_1}\xspace}
\newcommand{\ttwo   }{\ensuremath{\theta_2}\xspace}
\newcommand{\qzm    }{\ensuremath{\langle \qsq \rangle}\xspace}
\newcommand{\pzm    }{\ensuremath{\langle \psq \rangle}\xspace}
\newcommand{\pt     }{\ensuremath{p_{\rm t}}\xspace}
\newcommand{\restr  }{\ensuremath{{\sigma_{\pt}}/{\pt}=\sqrt{(0.02)^2+(0.0015\,\pt)^2}}\xspace}
\newcommand{\act    }{\ensuremath{|\cos\theta|}\xspace}
\newcommand{\cts    }{\ensuremath{\cos\theta^\star}\xspace}
\newcommand{\dsigdx}{\ensuremath{\mathrm{d}\sigma/\mathrm{d}x}\xspace}
\newcommand{\stt  }{\ensuremath{\sigma_\mathrm{TT}}\xspace}
\newcommand{\slt  }{\ensuremath{\sigma_\mathrm{LT}}\xspace}
\newcommand{\stl  }{\ensuremath{\sigma_\mathrm{TL}}\xspace}
\newcommand{\sll  }{\ensuremath{\sigma_\mathrm{LL}}\xspace}
\newcommand{\ttt  }{\ensuremath{\tau_\mathrm{TT}}\xspace}
\newcommand{\ttl  }{\ensuremath{\tau_\mathrm{TL}}\xspace}
\newcommand{\barph}{\ensuremath{\bar{\phi}}\xspace}
\newcommand{\cosph}{\ensuremath{\cos\barph}\xspace}
\newcommand{\costph}{\ensuremath{\cos 2\barph}\xspace}
\newcommand{\ropp }{\ensuremath{\rho_1^{++}}\xspace}
\newcommand{\rtpp }{\ensuremath{\rho_2^{++}}\xspace}
\newcommand{\ropm }{\ensuremath{\rho_1^{+-}}\xspace}
\newcommand{\rtpm }{\ensuremath{\rho_2^{+-}}\xspace}
\newcommand{\ronn }{\ensuremath{\rho_1^{00}}\xspace}
\newcommand{\rtnn }{\ensuremath{\rho_2^{00}}\xspace}
\newcommand{\ropn }{\ensuremath{\rho_1^{+0}}\xspace}
\newcommand{\rtpn }{\ensuremath{\rho_2^{+0}}\xspace}
\newcommand{\Ngt    }{\ensuremath{\frac{{\mathrm d}^2N_{\gamma}}{{\mathrm d}z{\mathrm d}\psq}}\xspace}
\newcommand{\Ngo    }{\ensuremath{\frac{{\mathrm d}N_{\gamma}}{{\mathrm d}z}}\xspace}
\newcommand{\me     }{\ensuremath{m_{\mathrm{e}}}\xspace}
\newcommand{\der    }{\ensuremath{{\mathrm d}}\xspace}
\newcommand{\faoft  }{\ensuremath{\faqed/\ftqed}\xspace}
\newcommand{\fboft  }{\ensuremath{\fbqed/\ftqed}\xspace}
\newcommand{\half   }{\ensuremath{\frac{1}{2}}\xspace}
\newcommand{\pz     }{\ensuremath{\phantom{0}}\xspace}
\newcommand{\prob   }{\ensuremath{P(\chi^2)}\xspace}
\begin{document}
 \begin{titlepage}
 \begin{center}{\large   EUROPEAN LABORATORY FOR PARTICLE PHYSICS
               }\end{center}\bigskip
 \begin{flushright}
        CERN-EP/99-010   \\ January 28, 1999
 \end{flushright}\bigskip\bigskip\bigskip\bigskip\bigskip
 \begin{center}
 {\huge\bf Measurements of the QED Structure of the Photon}
 \end{center}\bigskip\bigskip
 \begin{center}{\LARGE The OPAL Collaboration}\end{center}\bigskip\bigskip
 \bigskip\begin{center}{\large  Abstract}\end{center}
 The structure of both quasi-real and highly virtual photons
 is investigated using the reaction $\epem\rightarrow\epem\mupmum$,
 proceeding via the exchange of two photons.
 The results are based on the complete OPAL dataset taken at 
 \epem centre-of-mass energies close to the mass of the \zn boson.
 The QED structure function \ftqed and the differential cross-section 
 \dsigdx for quasi-real photons are obtained 
 as functions of the fractional momentum $x$ from the muon momentum
 which is carried by the struck muon in the quasi-real photon
 for values of \qsq ranging from 1.5 to 400 \gevsq. 
 The differential cross-section \dsigdx for highly virtual photons is 
 measured for $1.5< \qsq < 30$ \gevsq and $1.5< \psq < 20$ \gevsq, where
 \qsq and \psq are the negative values of the four-momentum
 squared of the two photons such that $\qsq>\psq$.
 Based on azimuthal correlations the QED structure functions \faqed and \fbqed 
 for quasi-real photons are determined for an average \qsq of 5.4 \gevsq.
\bigskip\bigskip\bigskip\bigskip
\bigskip\bigskip
\begin{center}
{\large (Submitted to Eur. Phys. Journal C)}
\end{center}
\end{titlepage}
\begin{center}{\Large        The OPAL Collaboration
}\end{center}\bigskip
\begin{center}{
G.\thinspace Abbiendi$^{  2}$,
K.\thinspace Ackerstaff$^{  8}$,
G.\thinspace Alexander$^{ 23}$,
J.\thinspace Allison$^{ 16}$,
N.\thinspace Altekamp$^{  5}$,
K.J.\thinspace Anderson$^{  9}$,
S.\thinspace Anderson$^{ 12}$,
S.\thinspace Arcelli$^{ 17}$,
S.\thinspace Asai$^{ 24}$,
S.F.\thinspace Ashby$^{  1}$
% ...
}\end{center}\bigskip
\bigskip
$^{  1}$School of Physics and Astronomy, University of Birmingham,
Birmingham B15 2TT, UK
\newline
$^{  2}$Dipartimento di Fisica dell' Universit\`a di Bologna and INFN,
I-40126 Bologna, Italy
\newline
% ...
$^{  5}$Cavendish Laboratory, Cambridge CB3 0HE, UK
\newline
% ...
$^{  8}$CERN, European Organisation for Particle Physics,
CH-1211 Geneva 23, Switzerland
\newline
$^{  9}$Enrico Fermi Institute and Department of Physics,
University of Chicago, Chicago IL 60637, USA
\newline
% ...
$^{ 12}$Indiana University, Department of Physics,
Swain Hall West 117, Bloomington IN 47405, USA
\newline
% ...
$^{ 16}$Department of Physics, Schuster Laboratory, The University,
Manchester M13 9PL, UK
\newline
$^{ 17}$Department of Physics, University of Maryland,
College Park, MD 20742, USA
\newline
% ...
$^{ 23}$Department of Physics and Astronomy, Tel Aviv University,
Tel Aviv 69978, Israel
\newline
$^{ 24}$International Centre for Elementary Particle Physics and
Department of Physics, University of Tokyo, Tokyo 113-0033, and
Kobe University, Kobe 657-8501, Japan
\newline
% ...
\clearpage
%
%
\section{Introduction}
\label{sec:intro}
\begin{figure}[h]\unitlength 1pt
\begin{center}
%
%
\begin{picture}(250,185)(0,0)
 \Line(0,150)(100,150)
 \Line(100,150)(200,180)
 \Photon(100,150)(150,120){4}{8.5}
 \Line(150,120)(150,80)
 \Line(150,120)(200,120)
 \Text(  0,155)[lb]{\large\bf e(p$_\mathbf 1$)}
 \Text(210,175)[lb]{\large\bf e$_{\mbox{tag}}$(p$_\mathbf 1^{\prime}$)}
 \Text(  0, 35)[lb]{\large\bf e(p$_\mathbf 2$)}
 \Text(210, 15)[lb]{\large\bf e$_{\mbox{stag}}$(p$_\mathbf 2^{\prime}$)}
 \Text(210,115)[lb]{\mbox{\boldmath\large $\mu$}}
 \Text(210, 75)[lb]{\mbox{\boldmath\large $\mu$}}
 \Text( 90,115)[lb]{\mbox{\boldmath\large $\gamma^{\star}(q)$}}
 \Text( 90, 80)[lb]{\mbox{\boldmath\large $\gamma^{(\star)}(p)$}}
 \Line(0,60)(80,60)
 \Line(80,60)(200,40)
 \Photon(80,60)(150,80){4}{10.5}
 \Line(150,80)(200,80)
\end{picture}
\caption{\label{fig:pr271_01}
          Diagram of the reaction 
          $\ee\rightarrow\ee\gmbg\rightarrow\ee\mupmum$. 
          The symbols in brackets denote the four-vectors of the particles
        }
\end{center}
\end{figure}
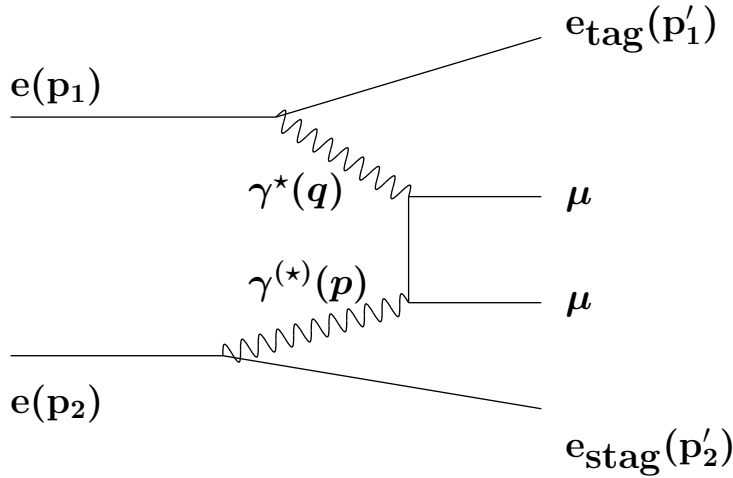

 The investigation of the structure of the photon represents a
 fundamental test of the predictions of QED and QCD.
 The classical method of investigation is the measurement of photon structure 
 functions in deep inelastic electron-photon scattering at \epem colliders.
 The photon couples to the electric charge and it reveals its
 structure in the fluctuations into virtual lepton and quark pairs.
 The pure QED process $\epem\rightarrow\epem\mupmum$, which 
 mainly proceeds via the exchange of two 
 photons, is an ideal environment free of QCD effects.
 In the phase space region under consideration the contribution of the 
 exchange of \zn bosons is negligible.
 For the largest part of the cross-section, both exchanged photons 
 are quasi-real 
 and the electrons\footnote{Electrons and positrons are referred to as
 electrons} are scattered, undetected, at small angles.
 If one of the photons is highly virtual the corresponding
 electron is usually scattered into 
 the acceptance of the detector and the reaction 
 ${\mathrm e}\gamma\rightarrow{\mathrm e}\gsg\rightarrow{\mathrm e} \mupmum$ 
 can be described as deep inelastic electron scattering off a quasi-real 
 photon, as illustrated in Figure~\ref{fig:pr271_01}.
 \par
%
%
\begin{figure}
\begin{center}
\mbox{\epsfig{file=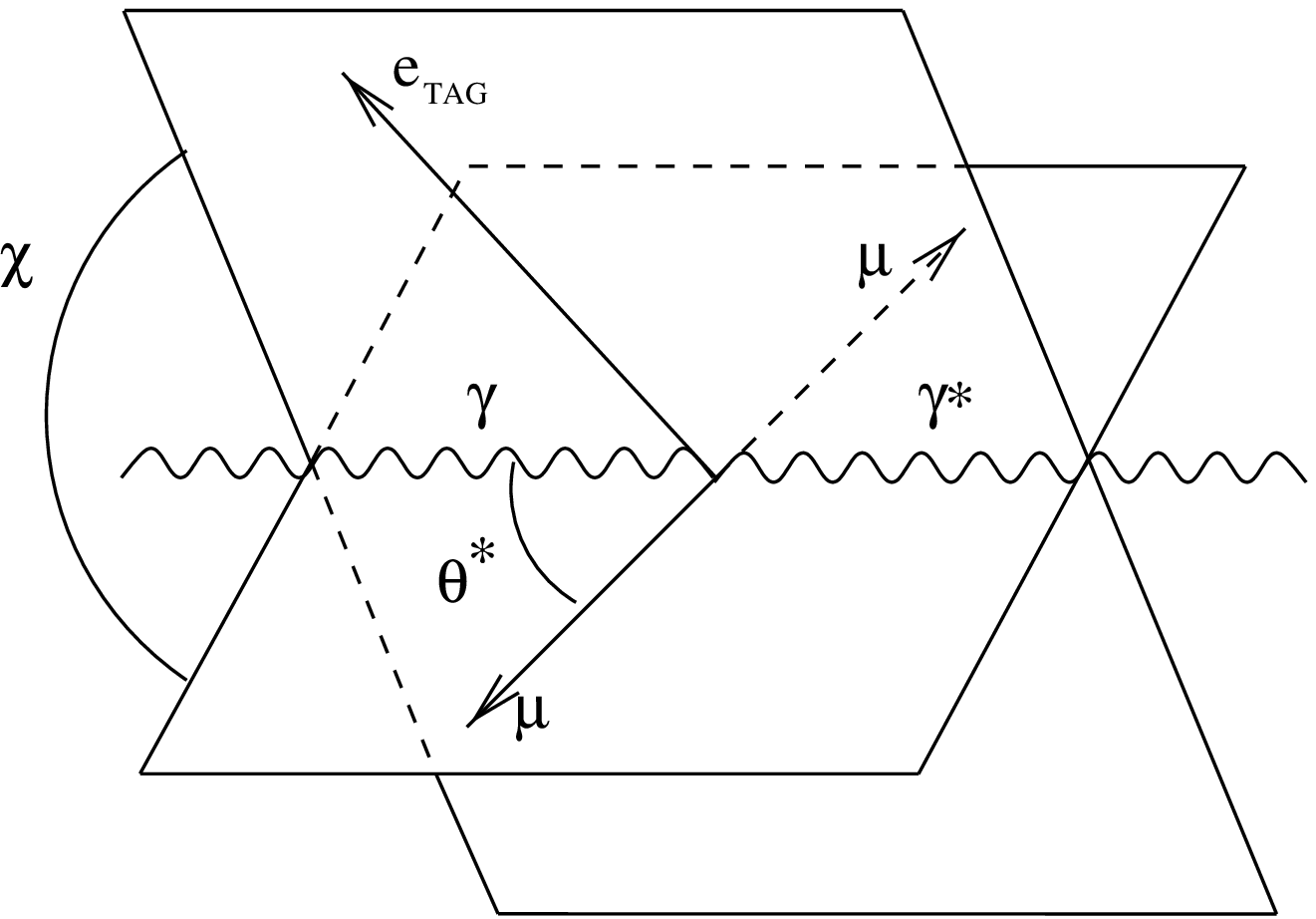,height=4.5cm}}
\caption{\label{fig:pr271_02}
  Illustration of the azimuthal angle $\chi$ for the reaction
  $\epem\rightarrow\epem\gsg\rightarrow\epem\mupmum$
  in the \gsg centre-of-mass system}
\end{center}
\end{figure}

%
%
 In this configuration the highly virtual photon, $\gamma^\star$, probes the 
 structure of the quasi-real photon, $\gamma$, and the
 structure functions of the  quasi-real photon can be measured.
 The differential cross-section~\cite{BUD-7501}
 \begin{equation}
  \frac{{\mathrm d}^2\sigma_{\rm e\gamma\rightarrow \rm e \mupmum}}
       {{\mathrm d}x{\mathrm d}Q^2}
 =\frac{2\pi\aemsq}{x\,Q^{4}}
  \left[\left( 1+(1-y)^2\right) \ftxqp - y^{2} \flxqp\right]
 \label{eqn:Xsect}
 \end{equation}
 at low values of $y$ is sensitive mainly to the QED structure function 
 \ftqed. 
 Here \qsq and \psq are the negative values of the four-momentum squared
 of the virtual photon and the quasi-real photon, respectively. 
 The symbols $x=\qsq/2p\cdot q$ and $y=p\cdot q / p_1\cdot p$ 
 denote the usual dimensionless variables of deep inelastic scattering, 
 and \aem is the fine structure constant.
 Due to the large statistics available at LEP, the dependence of \ftqed on the 
 small virtuality, \psq, of the quasi-real photon can be explored.
 The measurement of the distribution of the azimuthal angle, $\chi$,
 between the electron scattering plane and the
 plane containing the muon pair in the \gsg centre-of-mass system, 
 as defined in Figure~\ref{fig:pr271_02}, gives access to the structure 
 functions \faqed and \fbqed~\cite{PET-8301}, 
 as described in Section~\ref{sec:f2frame}.
%
%
%
%
 Real photons are only transversely polarised, but in the 
 configuration where both photons are highly virtual and both
 electrons are detected, the cross-section for the process 
 $\ee\rightarrow\ee\ggss\rightarrow\ee \mupmum$ 
 receives sizable contributions from longitudinal photons. 
 These contributions are large enough to be observed.
 \par
 Measurements of QED structure functions have been performed
 at several \epem colliders~\cite{CEL-8301TPC-8401PLU-8501}.
 Due to the clean final state, these measurements are limited mainly
 by statistics. 
 At LEP, the structure function \ftqed has been measured before
 by OPAL~\cite{OPALPR088} using a smaller data statistics, and by 
 DELPHI~\cite{DEL-9601} and L3~\cite{L3C-9801}.
 The distribution of the azimuthal angle was used to measure 
 the structure function ratio
 $\half\fboft$ by OPAL~\cite{OPALPR182} and \faqed and \fbqed
 by L3~\cite{L3C-9801}. 
 The differential cross-section \dsigdx of the reaction 
 $\ee\rightarrow\ee\mupmum$ mediated by 
 highly virtual photons has not been measured before.
 \par
 In the analysis presented here the full dataset of the OPAL detector 
 taken at LEP in the years 1990 $-$ 1995 at \epem centre-of-mass
 energies close to the mass of the \zn boson is used. 
 The structure function \ftqed and the differential cross-section 
 \dsigdx for quasi-real photons are extracted
 in the largest kinematic range ever covered by a single experiment. 
 In addition, the structure functions \faqed and \fbqed 
 for quasi-real photons are determined.
 For the first time a measurement of \dsigdx for highly virtual 
 photons is performed and the contributions of \ttl and \ttt to the 
 cross-section are established. Here \ttl and \ttt
 are interference terms which correspond to specific helicity states of
 the photons, as described in detail in Section~\ref{sec:f2frame}.
\par
 The paper is organised as follows.
 The theoretical framework is outlined in Section~\ref{sec:f2frame}.
 After a brief description of the OPAL detector in Section~\ref{sec:detec} 
 the kinematics and event selection are detailed in Section~\ref{sec:selec},
 followed by the discussion of the results in 
 Sections~\ref{sec:f2resu} and~\ref{sec:azresu}.
 Conclusions are drawn in Section~\ref{sec:concl}.
%
%
\section{Theoretical framework}
\label{sec:f2frame}
 In this section, the formalism used to extract the differential 
 cross-section and the structure functions is outlined. 
 For the measurement of \dsigdx and \ftqed, only the formulae integrated
 over the angular dependence of the \mupmum final state are relevant.
 On the other hand, the measurement of \faqed and \fbqed 
 involves the \az dependence of the \mupmum final state. 
 These two issues are discussed in turn. 
 \par
 The general form of the differential cross-section for the reaction 
 $\mathrm{e}(p_1) \mathrm{e}(p_2) \rightarrow \mathrm{e}(p_1^\prime) 
  \mathrm{e}(p_2^\prime) X$, which proceeds via the exchange
 of two photons $\gamma_1(q)$, $\gamma_2(p)$, 
 is given by
%
 \begin{eqnarray}
  \mathrm{d}^6\sigma 
  &=&
  \mathrm{d}^6\sigma(\ee\rightarrow\ee\,X)
  \nonumber
  \\
  &=&
  \frac{\mathrm{d^3p}_1^\prime\mathrm{d^3p}_2^\prime}{E_1'E_2'}
  \frac{\aemsq}{16\pi^4\sqsq\spsq}
  \left[
  \frac{(q\cdot p)^2-\sqsq\spsq}{(p_1\cdot p_2)^2
        -m_{\mathrm e}^2m_{\mathrm e}^2}
  \right]^{1/2}
  \nonumber
  \\
  &&
  \left(
  4\ropp\rtpp\stt +
  2 | \ropm\rtpm | \ttt\costph +
  2\ropp\rtnn\stl +
  2\ronn\rtpp\slt 
  \right.
  \nonumber
  \\
  &&
  \left.
  + \ronn\rtnn\sll -
  8 |\ropn\rtpn | \ttl\cosph
  \right)\, ,
 \label{eqn:true}
 \end{eqnarray}
%
 where $X$ denotes a fermion anti-fermion state.
 Here $p_1$ and $p_2$ represent the four-vectors of the incoming electrons,
 $p_1^\prime$ and $p_2^\prime$ the four-vectors of the scattered electrons
 and $q$ and $p$ the four-vectors of the exchanged photons
 $\gamma_1(q)$ and $\gamma_2(p)$, where $|\sqsq|>|\spsq|$ is chosen.
 The scattered electrons have energies $E_1'$ and $E_2'$, and
 \barph is the angle between the two scattering planes 
 of the electrons in the photon-photon centre-of-mass system.
 The cross-sections \stt, \stl, \slt and \sll
 and the interference terms \ttt and \ttl
 correspond to specific helicity states of the photons 
 (T=transverse and L=longitudinal)~\cite{BUD-7501}.
 Since a real photon can only have transverse polarisation, the terms where
 at least one photon has longitudinal polarisation have to vanish in the 
 corresponding limit $|\sqsq|\rightarrow 0$ or $|\spsq|\rightarrow 0$,
 and these terms have the following functional form:
 $\slt\propto\sqsq$,
 $\stl\propto\spsq$,
 $\sll\propto\sqsq\spsq$ 
 and
 $\ttl\propto\sqrt{\sqsq\spsq}$.
 The terms $\rho_1^{jk}$ and $\rho_2^{jk}$, 
 where $j,k\in(+,-,0)$ denote the photon helicities,
 are elements of the photon density matrix
 which only depend on the four-vectors $q$, $p$, $p_1$, $p_2$ and
 on the mass of the electron, $m_{\mathrm e}$. 
 They are listed in Ref.~\cite{BUD-7501}.
 \par
 In the case of muon pair production, $X=\mupmum$, the cross-section is 
 completely determined in QED.
 Equation~\ref{eqn:true} contains the full information, and it is
 sufficient to describe the reaction in terms of cross-sections.
 However, most of the experimental results are expressed 
 in terms of structure functions, since in the case of quark pair production 
 the cross-section cannot be completely calculated and has to be parametrised 
 by structure functions.
 The relations between the cross-sections and the structure functions 
 are given by~\cite{BER-8701} 
%
 \begin{eqnarray}
  2x\foqed &=&\frac{-\sqsq}{4\pi^2\aem}
           \frac{\sqrt{(q\cdot p)^2-\sqsq\spsq}}{q\cdot p}
           \left( \stt(x,\sqsq,\spsq) -\half\stl(x,\sqsq,\spsq)\right),
  \nonumber \\
  \ftqed &=& \frac{-\sqsq}{4\pi^2\aem} \,\,
          \frac{q\cdot p}{\sqrt{(q\cdot p)^2-\sqsq\spsq}} 
          \left(\frac{}{}
          \stt(x,\sqsq,\spsq) +\slt(x,\sqsq,\spsq)
          \right.
  \nonumber \\
         & & 
          \left.
          -\half\sll(x,\sqsq,\spsq) -\half\stl(x,\sqsq,\spsq)
          \right),
  \nonumber \\
  \flqed &=& \ftqed - 2x\foqed .
 \label{eqn:struc}
 \end{eqnarray}
%
 In the limit where one virtuality, e.g. $|\spsq|$, is small 
 and the other is large, $|\sqsq| \gg |\spsq|$, Eq.~\ref{eqn:true} 
 reduces to 
%
 \begin{equation}
 \frac{{\mathrm d}^4\sigma_{\ee\rightarrow\ee\mupmum}}
  {{\mathrm d}x {\mathrm d}\qsq {\mathrm d}z {\mathrm d}\psq}=
  \frac{2\pi\aemsq}{x\,Q^{4}}
  \cdot 
  \Ngt
  \cdot 
  \left[\left( 1+(1-y)^2\right) \ftxqp - 
  y^{2} \flxqp\right] ,
  \nonumber
 \label{eqn:approx}
 \end{equation}
%
 where $-\sqsq = \qsq$, $-\spsq = \psq$ and $z=E_\gamma/\eb$ is the 
 ratio of the energy of the quasi-real photon to the energy of the 
 beam electron radiating the quasi-real photon.
 In this formula the flux for quasi-real photons is expressed 
 using the equivalent photon approximation, EPA~\cite{KES-6001}:
%
 \begin{equation}
  \Ngt =  \frac{\aem}{2\pi}
           \left[
           \frac{1+(1-z)^2}{z}\frac{1}{\psq} -
           \frac{2\,\me^2\,z}{P^4}
            \right] .
%
 \label{eqn:oldpaper}
 \end{equation}
%
 For the experimental situation where the electron which radiates
 the quasi-real photon is not detected, the EPA is often used integrated 
 over the invisible part of the \psq range.
 The integration boundary \psqmin is given by four-momentum conservation 
 and \psqmax is determined by $\theta_\mathrm{max}$, the maximum angle 
 at which an electron carrying the energy \eb of the beam electrons
 could possibly escape detection.
 The integration of the EPA formula leads to the
 Weizs\"acker-Williams, WW, approximation~\cite{WEI-3401WIL-3401},
 which is a formula for the flux of collinear real photons:
%
 \begin{eqnarray}
  \Ngo &=& \int_{\psqmin}^{\psqmax}{\mathrm d}\psq\,\Ngt
 \nonumber \\
       &=&\frac{\aem}{2\pi}\left[
          \frac{1+(1-z)^2}{z}\ln\frac{\psqmax}{\psqmin}-
          2\,\me^2\,z\left(\frac{1}{\psqmin}-\frac{1}{\psqmax}\right)
          \right],\\
\mbox{where}\quad\psqmin &=& \frac{\me^2\,z^2}{1-z}, \quad\quad
\mbox{and}  \quad\psqmax  =  (1-z)\eb^2\theta^2_\mathrm{max}\nonumber.
 \label{eqn:weiz}
 \end{eqnarray}
%
 There are two potential problems with the approach of Eq.~\ref{eqn:approx}
 which are avoided in the analysis presented here by using Eq.~\ref{eqn:true}.
 Firstly, if only one photon is highly virtual and one electron is detected, 
 the use of the WW approximation is not adequate for the measurement of the
 \psq dependence of the structure function \ftqed for the quasi-real photon
 since then the dependences on \psq are inconsistently treated
 because the \psq dependence of the EPA is integrated 
 out, whereas the full dependence on \ftxqp on \psq is kept.
 In this case the EPA should be used. 
 Secondly, if both photons are highly virtual and both electrons are detected, 
 even the EPA is not applicable, since it is valid only for small values of
 \psq.
 \par
 If both photons are highly virtual, Eq.~\ref{eqn:true} can be evaluated
 in the limit $\qsq\gg m_{\mathrm e}^2$ and $\psq \gg m_{\mathrm e}^2$, 
 leading to
%
 \begin{eqnarray}
 \mathrm{d}^6\sigma 
  &=&
  \frac{\mathrm{d^3p}_1^\prime\mathrm{d^3p}_2^\prime}{E_1'E_2'}
  \frac{\aemsq}{16\pi^4\sqsq\spsq}
  \left[
  \frac{(q\cdot p)^2-\sqsq\spsq}{(p_1\cdot p_2)^2-
         m_{\mathrm e}^2m_{\mathrm e}^2}
  \right]^{1/2} 4\ropp\rtpp 
  \cdot 
  \nonumber
  \\
  &&
  \left(
  \stt + \stl + \slt + \sll + \half\ttt\costph -  4\ttl\cosph
  \right)\, .
 \label{eqn:truesimp}
 \end{eqnarray}
%
 If the interference terms \ttt and \ttl are independent of \barph, 
 the integration
 over \barph of the terms containing \cosph and \costph vanishes, and
 the cross-section is proportional to \stt + \stl + \slt + \sll. 
 In this case, Eq.~\ref{eqn:struc} can be used to define an
 effective structure function of virtual photons, as performed in
 Ref.~\cite{PLU-8405}.
 The total cross-sections and interference terms 
 formally depend only on \qsq, \psq, on the invariant mass squared, 
 \wsq, of the muon system and on the mass of the muon, $m_\mu^2$.
 However, there is a kinematical correlation between 
 these variables and \barph, which leads to the fact 
 that in several kinematical regions, like the one used in the analysis
 presented, \ttt and \ttl are not independent of \barph.
 Consequently, the terms proportional to \cosph and \costph do not vanish 
 even when integrated over the full range in \barph~\cite{ART-9501}.
 The resulting contributions can be very large, depending on the ratios
 $\qsq/\psq$, $\qsq/\wsq$ and $\psq/\wsq$. Numerical results of this effect
 can be found in Section~\ref{sec:f2resu}.
 Due to the large interference terms, cancellations occur between the 
 cross-section and interference terms and therefore no clear relation 
 of a structure function to the cross-section terms can be found.
 In this situation the cleanest experimentally accessible measurement
 is the differential cross-section \dsigdx as predicted
 by Eq.~\ref{eqn:truesimp}.
 \par
 The measurement of \faqed and \fbqed requires 
 the measurement of the \az distribution.
 Expressing the differential cross-section in terms which have 
 the same angular dependence with respect to the azimuthal angles 
 \az and \barph and combinations thereof,
 the differential cross-section can be written using
 13 structure functions as explained in Ref.~\cite{ART-9501}.
 By integrating over all angular dependences except the \az dependence,
 and factoring out the structure function \ftqed, only the structure function
 ratios \faoft and $\half\fboft$ remain for the deep inelastic electron
 photon scattering process and 
 these ratios can be obtained from a measurement of the \az distribution.
 For  deep inelastic electron photon scattering, the angle $\chi$ is defined 
 as the angle between the deeply inelastically scattered electron and 
 the muon which, in the 
 photon-photon centre-of-mass frame, is scattered at positive values of \cts, 
 as illustrated in Figure~\ref{fig:pr271_02}.
 To achieve sensitivity to the structure function \faqed
 the definition of \az is different from that used  
 in \protect Ref.~\cite{OPALPR182}. The old definition lead 
 to a vanishing term, proportional to $\cos\chi$, when 
 integrated of $\cos\theta^\star$. 
 With this new definition, the integration of Eq.~\ref{eqn:oldpaper} 
 of Ref.~\cite{OPALPR182} over \cts in the range $-$1 to 1 
 leads to
 \footnote{The definition of \faqed in Ref.~\cite{L3C-9801} differs by a 
 factor $-$1/2 from that used here, because in Ref.~\cite{L3C-9801}
 the angle \az is defined differently and the integration over \cts 
 is performed in the range 0 to 1 only}:
%
\begin{eqnarray}
\frac{\der\sigma (\mathrm{e}\gamma \rightarrow \mathrm{e}\mu^+\mu^-)}
     {\der x \der \qsq \der\chi/2\pi} 
   & =      & 
              \frac{2\pi\alpha^2}{xQ^4} \left(1+\left(1-y\right)^2\right)
              \times \nonumber \\
   &        &  {\ftqed} \left( 1 + \rho(y)(\faqed / \ftqed)\cos\chi +
               \frac{1}{2}\epsilon(y)( \fbqed / \ftqed ) \cos2\chi \right).
\label{eqn:xsect}
\end{eqnarray}
%
 Here $\rho(y)=(2-y)\sqrt{1-y}/(1+(1-y)^2)$ and
 $\epsilon(y)=2(1-y)/(1+(1-y)^2)$ as in Ref.~\cite{PET-8301}.
 Both are close to unity for small values of $y$. 
 Thus, \faoft and $\half\fboft$ are obtained from a fit to the \az
 distribution. By measuring \ftqed in addition, \faqed and 
 \fbqed can be calculated.
 Equation~\ref{eqn:xsect} is based on the structure functions for 
 real photons, $\psq=0$.
 The formulae for the structure functions \ftqed, \faqed and \fbqed
 are taken from Ref.~\cite{SEY-9801} and they keep the full
 dependence on the mass of the muon up to terms of order \msq/\wsq.
 The mass dependent formulae are significantly different, especially for
 the structure functions \faqed from the leading logarithmic approximation as, 
 for example, listed in Ref.~\cite{AUR-9601}.
 For example for $\qsq=5.4$~\gevsq and $x=0.8$,
 \faqed from Ref.~\cite{SEY-9801}
 is about 10$\%$ higher than \faqed using the 
 leading logarithmic form of Ref.~\cite{AUR-9601}.
 The functions, which are explicitly used for the reweighting procedure 
 explained in Section~\ref{sec:azresu}, have the following form:
%
\begin{eqnarray}
 \faqed(x,\beta) &=& 
 \frac{4\alpha}{\pi} x\sqrt{x\left(1-x\right)} \left(1-2x\right)
 \left\{
 \beta\left[1 + \left(1-\beta^2\right)\frac{1-x}{1-2x}\right]
 \right.\nonumber\\
                 & &
 \left.
 +\frac{3x-2}{1-2x}\sqrt{1-\beta^2}\arccos\left(\sqrt{1-\beta^2}\right)
 \right\},
 \label{eqn:azfu1}\\
 \fbqed(x,\beta) &=& 
 \frac{4\alpha}{\pi}x^2\left(1-x\right)
 \left\{\beta\left[1-\left(1-\beta^2\right)\frac{1-x}{2x}\right] 
 \right.\nonumber\\
                 & &
 \left.
 +\frac{1}{2}\left(1-\beta^2\right)
 \left[\frac{1-2x}{x}-\frac{1-x}{2x}\left(1-\beta^2\right)\right]
 \log\left(\frac{1+\beta}{1-\beta}\right)\right\},\\
 \ftqed(x,\beta) &=& 
 \frac{\alpha}{\pi} x\left\{
 \left[x^2+\left(1-x\right)^2\right]\log\left(\frac{1+\beta}{1-\beta}\right)
 -\beta 
 +8\beta x\left(1-x\right)
 \right.
 \nonumber\\&&
 \left.
 -\beta\left(1-\beta^2\right)\left(1-x\right)^2
 \right.
 \nonumber\\&&
 \left.
 +\left(1-\beta^2\right)\left(1-x\right)
 \left[\frac{1}{2}\left(1-x\right)\left(1+\beta^2\right)-2x\right]
 \log\left(\frac{1+\beta}{1-\beta}\right)
 \right\}
 \label{eqn:azfu2},\\
 \mbox{where } \beta           &=& \sqrt{1-\frac{4m^2_{\mu}}{\wsq}}\nonumber .
\end{eqnarray}
%
 \par
 The QED cross-section, Eq.~\ref{eqn:true}, 
 which keeps the full dependence on the virtualities of both photons,
 is implemented in the Monte Carlo programs 
 Vermaseren~\cite{SMI-7701VER-7901VER-8301,BHA-7701},
 BDK~\cite{BDK-8501BDK-8601BDK-8602BDK-8603} and
 GALUGA~\cite{SCH-9801}.
 In the implementations used here, all programs only contain the 
 multiperipheral diagram shown in Figure~\ref{fig:pr271_01}.
 The contributions from the bremsstrahlung processes~\cite{BHA-7701}
 are small at low \qsq and get more important as \qsq increases.
 In this analysis the bremsstrahlung processes
 are treated as background using the predictions of the 
 FERMISV~\cite{HIL-9301} and the grc4f~\cite{FUJ-9801} programs, and
 possible interferences are neglected.
 The BDK program, in addition, contains radiative corrections.
 In the analysis presented here the Vermaseren program 
 is used to generate a large size
 event sample which is fully simulated and treated like the data. 
 The BDK program is used to determine the radiative corrections and
 the GALUGA program is used to calculate the individual contributions to 
 the cross-section in Eq.~\ref{eqn:true}.
 \par
 In the analysis presented here, the measurement of the differential 
 cross-section \dsigdx for events where one electron is observed 
 (singly-tagged events) is compared to the QED prediction 
 of Eq.~\ref{eqn:true}. 
 The interpretation in terms of the structure function \ftqed uses 
 the relation between the cross-sections and the structure functions,
 Eq.~\ref{eqn:struc}.
 The measurement of the differential cross-section \dsigdx for
 events where both electrons are observed (doubly-tagged events) is compared
 to the QED predictions using Eq.~\ref{eqn:truesimp}.
 The structure functions \faqed and \fbqed are obtained
 from a fit to the $\chi$ distribution using Eq.~\ref{eqn:xsect}.
%
%
\section{The OPAL detector}
\label{sec:detec}
 The OPAL detector is described in detail
 elsewhere~\cite{OPALPR021ALL-9301ALL-9401AND-9401}.
 Here only the subdetectors which are most relevant for this analysis
 are briefly discussed.
 In the OPAL right-handed coordinate system the $x$-axis points towards the
 centre of the LEP ring, the $y$-axis points upwards and the $z$-axis points in
 the direction of the electron beam. The polar angle \thet and the
 azimuthal angle \ph are defined with respect to the $z$-axis
 and $x$-axis respectively.
 \par
 The OPAL detector has a uniform magnetic field of 0.435~T
 along the beam direction throughout the central tracking
 region, with electromagnetic and hadronic calorimetry and muon chambers
 outside the coil.
 The small-angle silicon tungsten calorimeter (SW)
 covers the region in \thet from 25 to 59~mrad
 at each end of the OPAL detector.
 The unobstructed acceptance of the forward detectors (FD) covers the 
 \thet region from 60 to 140~mrad at each end of the OPAL detector.
 Both ends of the OPAL detector are equipped with
 electromagnetic endcap calorimeters (EE) covering the 
  polar angle range from 200 to 630~mrad on each side. 
 Charged particles are detected by a silicon microvertex detector, a drift
 chamber vertex detector, a large volume jet chamber and a set of
 $z$-chambers.
 The resolution of the transverse momentum for charged particles is \restr
 for central tracks, where \pt is in \gev, and degrades for higher 
 values of \act.
 The magnet return yoke is instrumented with streamer tubes
 for hadron calorimetry and
 is surrounded by several layers of muon chambers.
%
%
\section{Kinematics and data selection}
\label{sec:selec}
 The observed electron which radiated the photon of higher virtuality 
 is denoted as {\rm 'tag'} and, and in doubly-tagged events, the second
 observed electron is called {\rm 'stag'}.
 The virtualities of the photons, \qsq and \psq with $\qsq>\psq$,
 are determined from the energies, \etag and \estag, and polar angles, 
 \ttag and \tstag, of the detected electrons using the relations
 $\qsq = 2\,\eb\,\etag\,(1- \cos\ttag)$
 and
 $\psq = 2\,\eb\,\estag\,(1- \cos\tstag)$,
 where \eb denotes the energy of the beam electrons.
 By measuring the two electrons and the two muons, 
 or by assuming that the second electron is scattered at $\cos\tstag=\pm 1$
 in the case of singly-tagged events, the kinematics is constrained.
 This constrain is used to improve on the 
 calorimetric energy measurement of the electrons.
 Using conservation of energy and longitudinal momentum, the relation
\begin{equation}
\etag  = \frac{p_{\mu^+\mu^-}\cos\theta_{\mu^+\mu^-} +
               (2 \eb - E_{\mu^+\mu^-}) \cos\tstag}
               {\cos\tstag - \cos\ttag}
\label{eqn:ecalcul}
\end{equation}
 is derived. This formula allows to calculate the electron energies 
 from their polar angles, and the energy $E_{\mu^+\mu^-}$,
 the momentum $p_{\mu^+\mu^-}$ and polar angle $\cos\theta_{\mu^+\mu^-}$
 of the muon pair.
 The dimensionless variables of deep inelastic scattering are calculated
 using:
%
  \begin{equation}
   y = 1-\frac{\etag}{2\eb}\left(1+\cos\ttag\right)\, ,
   \quad\quad\quad\quad
   x = \frac{\qsq}{\qsq+\wsq+\psq}\, .
  \label{eqn:Xcalc}
  \end{equation}
%
 For the singly-tagged events, \psq is much smaller
 than \qsq and is therefore neglected in the determination of $x$.
 \par
 The event selection requires in addition to one or two electrons of high 
 energy detected in the electromagnetic calorimeters SW, FD or EE, the 
 presence of exactly two charged particles with opposite charge which are 
 not associated with the energy clusters of the scattered electrons. 
 At least one of the charged particles has to be identified as a muon.
 The samples of singly-tagged events are denoted hereafter by the abbreviation
 of the calorimeter in which the scattered electron is detected, SW, FD or EE. 
 The samples of doubly-tagged events are denoted DB. 
 If a sample is further subdivided 
 in \qsq and \psq, the subsamples are called, for example, SW1 or FD2.
 The samples are defined in Table~\ref{tab:samp}.
 For singly-tagged events, all events are vetoed which contain 
 electromagnetic clusters with an energy larger than a certain fraction of 
 \eb in the hemisphere opposite to the one containing the observed electron. 
 For the SW, FD and EE samples the different background contributions
 lead to energy fraction cuts of 20$\%$, 20$\%$ and 5$\%$, respectively. 
 \par
 Electromagnetic clusters are accepted as electrons if they fulfill
 the following criteria:
%
 \begin{Enumerate}
 \item The energy of the cluster is larger than half the energy of the
       beam electrons.
 \item The polar angle $\theta$ of the cluster is in the range 
       $28 - 55$~mrad (SW),
       $60 - 120$~mrad (FD) or $210 - 540$~mrad (EE) with respect to either
       beam direction, respectively. The numbers are chosen such that the 
       electrons are well contained in the calorimeters. 
       As the cross-section for the signal events falls off more steeply with
       $\theta$ than the cross-section for the bremsstrahlung processes,
       the upper limit of $\theta$ for the EE sample is taken to be $540$~mrad.
       This is a tighter limit than that required for good containment, 
       but is applied to the EE sample to ensure a high signal to background 
       ratio.
 \end{Enumerate}
%
 A track is accepted as a charged particle if it satisfies the 
 following criteria:
%
\begin{Enumerate}
 \item It has at least 20 hits in the jet chamber.
 \item The distance of the point of closest approach to the 
       origin in the $r\phi$-plane is less than 1.0~cm in the 
       $r\phi$-plane and less than $20$~cm in the $z$-direction.
 \item The momentum is between 0.3 and 20 \gev, and the
       transverse momentum relative to the $z$-direction is greater 
       than 0.1 \gev.
 \item The polar angle of the track is within the clean acceptance 
       of the OPAL jet chamber which extends to $\act = 0.96$.
\end{Enumerate}
%
 A particle is identified as a muon if it meets the following criteria:
%
 \begin{Enumerate}
 \item The momentum is greater than 1 \gev.
 \item The energy deposit in the electromagnetic calorimeter 
       associated with the track is less than 1.5 \gev.
 \item It produces a muon signal either in the hadron calorimeter 
       or in the muon chambers, as described in Ref.~\cite{OPALPR088}.
 \end{Enumerate}
%
 The SW, FD, EE and DB samples have different kinematical distributions for
 the muons and electrons and different contributions from background 
 processes.
 Therefore, the samples have to fulfill different trigger conditions and
 some specific cuts are applied in addition to those previously described.
%
 \begin{Enumerate}
 \item \underline{SW sample:}\\
       In order to ensure a trigger efficiency close to 100$\%$,
       which can reliably be estimated from the data,
       it is required that at least one muon is observed in the region
       $|\cos \theta_{\mu}| \le 0.74$.
       This restricted range in $|\cos \theta_{\mu}|$ however
       is only required for the measurement of \ftqed and \dsigdx and 
       not when measuring the $\chi$ distribution to obtain 
       \faoft and $\half\fboft$.
       This choice is made because the accuracy of the measurement of \faoft 
       and $\half\fboft$ is limited by the statistical error, and the 
       measurement only relies on event ratios and does not use absolute 
       cross-sections. Therefore in order to retain the highest possible 
       number of events all muons up to $|\cos\theta_{\mu}|=0.96$ are
       accepted for the measurement of \faoft and $\half\fboft$.
 \item \underline{EE sample:} \\
       A large background comes from \znmu events with a high-energy
       photon radiated by one of the muons and where this photon is
       identified as an electron.
       As these photons tend to be close to the muons, this background
       is effectively rejected by requiring that the distance
       $R=\sqrt{(\Delta\eta)^2+(\Delta\phi)^2}$
       between the momentum vector of the electron candidate and the 
       momentum vector of the closest muon is larger than unity.
       Here $\eta=-\ln(\tan(\theta/2))$ is the pseudorapidity with respect 
       to the $z$-axis and $\phi$ the azimuthal angle.
       In addition, at least one muon with momentum larger than
       5 \gev and a minimum invariant mass squared, $\wsq > 1$ \gevsq,
       is required.
 \item \underline{DB sample:}\\
       Due to the requirement of two observed electrons, the background 
       is much reduced. The isolation requirement is loosened to
       $R \ge 0.5$ with respect to the electron candidates in the EE
       calorimeter.
 \end{Enumerate}
%
 The $\ee\rightarrow\ee\mupmum$ events are triggered with high efficiency 
 by the large energy deposits of the scattered electrons in the 
 electromagnetic calorimeters 
 and by the muons detected in the tracking devices and muon chambers.
 The trigger efficiency for events fulfilling all kinematical cuts 
 is determined from the data using triggers which are either related to the 
 scattered electrons or to the muon tracks.
 The trigger efficiency is studied separately for the individual samples.
 For the EE and the FD samples completely independent triggers for electrons
 and muons are available and the trigger efficiencies are found to be 
 100$\%$ and 98$\%$, respectively, and do not vary with $x$.
 For the SW sample the situation is more complicated
 because no trigger is available which is only based on the electromagnetic 
 cluster produced by the electron. 
 Therefore the acceptance in polar angle for at least one of the muons 
 has to be reduced to get a reliable trigger estimate,
 especially for large values of $x$ which correspond to low invariant
 masses of the muon system.
 The trigger efficiency for the SW sample is found to be
 98$\%$ to 99$\%$ and does not vary with $x$ in the ranges of $x$ used
 for the analysis, as listed in Table~\ref{tab:samp}.
 The error on the trigger efficiencies are conservatively taken as the 
 difference of the evaluated trigger efficiencies from 100$\%$.
 \par
 Due to different detector configurations and status requirements,
 the available luminosities for the various samples are different.
 The luminosities used amount to 67.4 \invpb for the SW and DB 
 samples, to 81.3 \invpb for the FD sample,
 and to 129.1 \invpb for the EE sample.
 The number of events observed, together with the Monte Carlo predictions,
 are given in Table~\ref{tab:samp}.
 The integrated luminosities of the signal events, simulated with the 
 Vermaseren program, amount to 336 \invpb for the SW and FD samples, 
 919 \invpb for DB samples and 964 \invpb for the EE sample.
 It is verified that the cross-sections obtained with the 
 Vermaseren~\cite{SMI-7701VER-7901VER-8301,BHA-7701} 
 and GALUGA~\cite{SCH-9801} 
 programs agree with each other to within 1$\%$ for all samples.
 The background processes considered are 
 \ggtau and \gghad, based on the Vermaseren and 
 HERWIG~\cite{MAR-8801-KNO-8801-CAT-9101-ABB-9001-SEY-9201}
 generators,
 \zntau, \znmu, \zne simulated with 
 KORALZ~\cite{JDH-9401},
 and all processes with $\gamma$ or \zn boson exchange containing four 
 fermions in the final states as predicted by grc4f~\cite{FUJ-9801} 
 and FERMISV~\cite{HIL-9301}.
 The by far dominant source of background is the process
 \ggtau in all samples. For the EE sample the reaction \zntau is of similar 
 importance as the reaction \ggtau.
 All Monte Carlo events are passed through the GEANT simulation of the 
 OPAL detector~\cite{ALL-9201} and are subject to the same analysis
 as the OPAL data.
%
%
\section{Results for \boldmath \ftqed \unboldmath
                    and \boldmath \dsigdx \unboldmath}
\label{sec:f2resu}
 The differential cross-sections \dsigdx and the structure functions
 \ftqed are unfolded from the observed distribution in $x$ of the data
  by means of a regularised unfolding technique~\cite{BLO-8401BLO-9601}.
 The data from the different samples are analysed separately. 
 For the unfolding of the structure function \ftxqp,
 the \qzm and \pzm values listed in Table~\ref{tab:samp} are used. 
 The average values of \qzm and \pzm for the different samples as predicted 
 by the Monte Carlo agree well with the values observed in the data.
 The value for \pzm for the SW, FD and EE samples is taken from the 
 Monte Carlo.
 The signal definition is based on the multipheripheral 
 diagram only and the bremsstrahlung diagrams are treated as background.
 The contribution of the bremsstrahlung diagrams to the cross-section 
 is less than 0.5$\%$ for the SW and FD sample and approximately 1.6$\%$ 
 for the EE sample.
 The resolution in \qsq is determined from the signal Monte Carlo.
 The resolution increases with increasing \qsq and ranges from 
 about 2.3$\%$ for electrons detected in the SW calorimeter to about 3.2$\%$ 
 for electrons detected in the EE calorimeter.
 \par
 For each of the samples, the agreement between the data and the
 Monte Carlo predictions, for signal and background events, is checked 
 by comparing quantities reconstructed from the electrons and the muons.
 The variables used are the energies \etag, \estag, \eone, \etwo,
 and angles \ttag, \tstag, \tone, \ttwo, 
 of the detected electrons and of the first (1) and second (2) muon,
 and the derived quantities $y$, \qsq, \psq and $W$.
 Good agreement of the distributions is found both in shape and normalisation 
 for all samples.
 The agreement between the data and the Monte Carlo predictions
 of the SW and FD samples is similar to the findings of Ref.~\cite{OPALPR182}.
 Some examples of control distributions for the EE sample and
 the DB samples are shown in 
 Figures~\ref{fig:pr271_03} and~\ref{fig:pr271_04}.
 The data with their statistical errors are compared to the signal 
 Monte Carlo with the background added to it before the unfolding 
 is performed, and to the Monte Carlo distributions after 
 reweighting the signal events, where the weights are obtained 
 from the unfolding procedure.
 All distributions show a good description of the data 
 by the sum of signal and background Monte Carlo events. 
 Figure~\ref{fig:pr271_05} shows a comparison between the data
 and the Monte Carlo prediction for the DB samples before
 the unfolding using statistical errors only.
 Shown are the measured azimuthal angle \barph between the two 
 scattering planes 
 of the electrons in the photon-photon centre-of-mass system
 and the measured azimuthal angle, $\chi$, which, for doubly-tagged events,
 is defined with respect to the plane containing the electron which radiated
 the photon of higher virtuality.
 Both angular distributions of the data are well reproduceded by the Vermaseren
 program. The data exhibit a strong dependence 
 on \barph which is well described by the prediction of Eq.~\ref{eqn:true}.
 These distributions in principle give access to several other structure 
 functions~\cite{ART-9501} but, due to the low statistics, no detailed
 analysis of these distributions has been performed.
 \par
 The differential cross-sections \dsigdx and the structure functions \ftqed 
 are unfolded from the observed $x$ distribution of the data, shown 
 in Figure~\ref{fig:pr271_06}.
 The distributions in Figure~\ref{fig:pr271_06} 
 are not corrected for radiative effects and trigger inefficiencies.
 Because the predicted $x$ distributions are close to the observed 
 distributions, the results of the unfolding are also very close to
 the QED prediction.
 The method used to unfold the cross-section \dsigdx and
 the structure function \ftqed closely follows the procedure 
 applied to the measurement of the hadronic structure function \ft and is
 described in detail in Ref.~\cite{OPALPR185OPALPR207OPALPR213}. 
 There is however a very important difference between
 the \mupmum final state and the hadronic final state.
 The resolution in $x$ of the \mupmum final state which, 
 as determined form the signal Monte Carlo, amounts to about 0.03,
 is much better than the resolution in $x$ for the hadronic final state
 due to the good measurement of the invariant mass of the muon pair.
 Therefore a much finer binning in $x$ with much reduced correlations
 between the bins, could be chosen here.
 \par
 The differential cross-sections \dsigdx and the structure functions 
 \ftqed, normalised by the fine structure constant \aem,
 unfolded from the data, are shown in Figure~\ref{fig:pr271_07} 
 and Figure~\ref{fig:pr271_09} for the independent samples 
 of singly-tagged events and in Figure~\ref{fig:pr271_08} 
 and Figure~\ref{fig:pr271_10} for the combined SW and FD samples.
 The measured values are corrected both for radiative effects and for the
 trigger inefficiencies.
 The radiative corrections are based on the ratio of the
 predicted cross-sections by the BDK and Vermaseren programs
 in bins of $x$. The radiative corrections vary with \qsq and $x$ and 
 amount up to about $\pm 9\%$.
 The structure function values are given at the centres of the $x$ bins.
 Because, in a given bin in $x$, the average value of \ftqed is
 different from the value of \ftqed at the centre of the $x$ bin,
 the results has to be corrected for this bin size effect.
 The measured average value of \ftqed in a given bin in $x$, as obtained 
 from the unfolding, is corrected for the bin size effect
 by multiplying the measured value of \ftqed with the QED prediction of the
 ratio of \ftqed at the centre of the bin and the average \ftqed in the bin.
 In general the corrections are small and the largest corrections occur 
 at low, and high values of $x$. The corrections are below 1$\%$ in all but
 the lowest and highest bins in $x$, for all samples. 
 In the lowest and highest bins in $x$ the correction is always positive, and
 amounts up to 7$\%$ in the lowest $x$ bins, and in the highest $x$ bins 
 it is around 5$\%$, with the exception of the SW1 sample, where it is largest
 and amounts to 19$\%$.
 The vertical error bars show both the statistical error and the full
 error, which is obtained from the quadratic sum of statistical and systematic 
 errors.
 In all unfolded results, the statistical error is obtained from the quadratic
 sum of the statistical error of the data events and 
 the signal Monte Carlo events.
 The measured values for the cross-sections \dsigdx are listed
 in Tables~\ref{tab:sf2sw}$-$\ref{tab:sf2db} and 
 the structure functions in Tables~\ref{tab:ff2sw}$-$\ref{tab:ff2ee}. 
 For the cross-section measurement of the singly-tagged events, 
 the corrected data correspond to the phase space defined by $y<0.5$, 
 $\psq<1.4$~\gevsq and the \qsq ranges listed in Table~\ref{tab:samp}.  
 The full range of $W$ is used, except for the EE sample where 
 $\wsq>1$~\gevsq is required.
 For the doubly-tagged events, 
 the corrected data correspond to the phase space defined by $y<0.5$ 
 and the \qsq and \psq ranges listed in Table~\ref{tab:samp}.
 \par
 The systematic error receives contributions from several sources.
 The determination of $x$ and \qsq is based on 
 the measurement of the muon and electron momenta.
 The uncertainties in these measurements are taken into account 
 by shifting the reconstructed quantities in the Monte Carlo samples
 according to resolution and repeating the unfolding. 
 The variations performed are the following.
%
 \begin{Enumerate}
 \item The transverse momentum of the muons is shifted by 0.25$\%$ 
       for tracks in the region $|\cos\theta_{\mu}|<0.90$ and by
       5$\%$ for very forward tracks, which are less well measured up to 
       $|\cos\theta_{\mu}|=0.96$.  
 \item The polar angle and the azimuthal angle of the muons are shifted
       by 0.2~mrad for tracks in the region $|\cos\theta_{\mu}|<0.90$ and by
       1~mrad for very forward tracks up to $|\cos\theta_{\mu}|=0.96$.  
 \item The polar angle of the observed electron is shifted by 0.3~mrad,
       0.7~mrad and 5~mrad for electrons observed in the SW, FD and EE 
       calorimeters, respectively.
 \end{Enumerate}
%
 The differences of the results based on the central values and the results 
 obtained using the shifted values are added in quadrature. 
 The systematic errors due to the uncertainties in the
 determination of trigger efficiencies and of the radiative corrections
 are also added in quadrature. The uncertainty is dominated in almost all 
 bins by the statistical error.
 \par
 The predicted structure function \ftqed is strongly suppressed for
 $\psq=\pzm$ compared to $\psq=0$.
 The measured structure functions Figures~\ref{fig:pr271_09} 
 and~\ref{fig:pr271_10} are distinctly 
 different from the predictions for $\psq=0$ for all values of \qsq.
 In general there is good agreement between the data 
 and the predictions for all ranges in \qsq, and the corresponding $\chi^2$ 
 probabilities \prob are listed in Tables~\ref{tab:sf2sw}$-$\ref{tab:ff2ee}.
 Some small differences between the data and the predictions can be seen for
 $\qsq > 10$~\gevsq. For the individual FD samples, FD2 and FD3,
 the data shows a slightly different shape for $x<0.5$ than the predicted 
 differential cross section, Figure~\ref{fig:pr271_07}(d,e), and the
 structure function, Figure~\ref{fig:pr271_09}(d,e), and 
 for the EE sample the predicted differential cross section,
 Figure~\ref{fig:pr271_07}(f), and the structure function 
 Figure~\ref{fig:pr271_09}(f),
 are slightly higher than what is observed in the data for all $x$.
 \par
 The cross-sections for the samples DB1 and DB2,
 unfolded from the data, are listed in Table~\ref{tab:sf2db} and
 shown in Figure~\ref{fig:pr271_11}, together
 with the predictions from the Vermaseren and GALUGA programs.
 The data are well described by both Monte Carlos using the full 
 cross-section from Eq.~\ref{eqn:true}. 
 Using the GALUGA predictions,
 the influence of the non-vanishing terms proportional to \cosph
 and \costph can be seen. If these terms are neglected,
 the predicted cross-section grossly overestimates the measured 
 cross-section.
 This shows that both terms, \ttt and \ttl, are present,
 mainly at $x>0.1$, and that the corresponding contributions to the 
 cross-section are negative. 
 The contribution from \ttl is especially very large 
 in the specific kinematical region of the DB samples.
 There is also good agreement between the number of events predicted and 
 observed for the DB3 sample but, because of its low statistics,
 no cross-section is evaluated for that sample. 
%
%
\section{Results for \boldmath $\bf\faqed$ \unboldmath and
                     \boldmath $\bf\fbqed$ \unboldmath}
\label{sec:azresu}
 For the combined SW and FD samples,
 the quantities \faoft and $\half\fboft$ are obtained in bins of $x$ 
 from the azimuthal angle distributions shown in Figure~\ref{fig:pr271_12}.
 The resolution in $\chi$ is about 20~mrad over the whole range 
 of $\chi$. The measured distributions do not exhibit directly the 
 $\cos\chi$ and $\cos 2\chi$ dependences predicted by QED
 in Eq.~\ref{eqn:xsect}.
 This is mainly due to the loss of muons close to the beam axis
 in the region $|\cos\theta_\mu|>0.96$.
 In order to extract \faoft and $\half\fboft$, the 
 azimuthal angle distributions are corrected for this and other detector
 effects using a bin-by-bin correction in $\chi$ for ranges of $x$.  
 The correction factor is a given bin is obtained from
 the distribution of the generated $\chi$ for events passing the selection 
 cuts, where each event is weighted by 
 $1/(1+\faoft\cos\chi+\frac{1}{2}\fboft\cos 2\chi)$.
 The ratios \faoft and \fboft used in the weighting function 
 are determined from the analytical QED structure 
 functions Eq's.~\ref{eqn:azfu1}$-$\ref{eqn:azfu2}. 
 The weighted distribution is equivalent to what 
 would be obtained from a flat \az distribution.
 The corrected distributions are shown in Figure~\ref{fig:pr271_13}.
 They are fitted to the following function:
%
 \begin{equation}
 F(\chi) = A (1 + B\cos\chi + C\cos 2\chi),
 \label{eqn:fit}
 \end{equation}
%
 where $A$ is a normalisation factor, which is left free in the fit,
 $B=\faoft$, and $C=\half\fboft$, for $\epsilon=\rho=1$.  
 The correlation coefficients between the parameters $B$ and $C$
 resulting from the fits are $0.04$ over the whole $x$ range and
 $0.26$, $0.13$, $-0.11$ and $-0.29$ for the different bins in $x$, as given
 in Table~\ref{tab:azre}.
 The measured average values of \faoft and $\half\fboft$ 
 in bins of $x$, as obtained from the fit, are corrected for the bin size 
 effect as before. 
 The measured values of $\half\fboft$ are in good agreement with the previous
 OPAL measurement of $\half\fboft$~\cite{OPALPR182}.
 The corrected values of \faoft and $\half\fboft$ 
 are listed in Table~\ref{tab:azre} and shown in 
 Figure~\ref{fig:pr271_14} compared to the QED predictions for 
 $\qsq = 5.4$~\gevsq and $\epsilon=\rho=1$. 
 The QED predictions for the full range in $x$ of $\faoft = -0.027$
 and $\half\fboft = 0.078$ are in good agreement with
 the measured values, $\faoft = -0.036\pm 0.027\pm 0.004$ and 
 $\half\fboft = 0.061\pm 0.013\pm 0.004$.
 The measured values of $\half\fboft$ as a function of $x$ are significantly
 different from zero and the measured shape agrees with the QED prediction, 
 although it is not significantly different from a constant.
 \par
 The sources of systematic errors are 
%
 \begin{Enumerate}
 \item \underline{Detector effects}:\\
       The same variations as carried out for the measurement of \ftqed and 
       \dsigdx and described in Section~\ref{sec:f2resu}, have been performed.
 \item \underline{Reweighting procedure}:\\  
       The reweighting procedure is tested by using a version of the 
       TWOGEN~\cite{BUI-9401} generator where the structure functions 
       \faqed and \fbqed can be chosen arbitrarily. 
       By chosing one of the two structure functions to be a constant
       and measuring the other structure function, systematic 
       uncertainties varying between 0.03$\%$ and 3.3$\%$ for the 
       measurement of \faoft and between 0.7$\%$ and 3.8$\%$ for 
       $\half\fboft$ in the different $x$ ranges are found.
 \item \underline{Background}: \\
       As the background is not subtracted from the data, the fit to the 
       data is a superposition of the fit to the signal and background 
       distributions contributing with their relative fractions.
       To take the effect of the background distribution into account
       in the estimation of the systematic error, 
       the $\chi$ distribution of the background is corrected,
       using the same procedure as for the data, and then fitted 
       like the data. 
       The parameters $B$ and $C$ of this fit are weighted by the relative 
       contribution of the background to the data for each range in $x$.
       They constitute the estimate of the systematic error, which
       varies from 1.4$\%$ to 4.2$\%$ between the low and high $x$ range.
\end{Enumerate}
%
 The strength of the \az dependence varies with the scattering 
 angle $\cts$ of the muons in the photon-photon centre-of-mass system.
 Reducing the acceptance of $\cts$ enhances the \az 
 dependence but, to obtain a result for \faqed and \fbqed which is
 valid for the full range of $\cts$ the measurement has to be extrapolated
 using the predictions of QED.
 The measurements of Ref.~\cite{L3C-9801} are obtained in the range 
 $|\cts|<0.7$, and extrapolated to the full range in \cts, whereas 
 the measurement presented here is valid for the full angular range $|\cts|<1$.
 Taking this into account,
 the results for \faoft and $\half\fboft$ obtained here and the 
 measurements of Ref.~\cite{L3C-9801} are consistent.
 \par
 From the measurements of \faoft and $\half\fboft$ and \ftqed, the values 
 of \faqed and \fbqed are obtained in the following way.
 The data from the SW and FD sample are combined and the 
 structure function \ftqed is unfolded using the same bins 
 as those used for the measurement of \faoft and $\half\fboft$.
 As Eq's.~\ref{eqn:azfu1}$-$\ref{eqn:azfu2}, used for the measurement of 
 \faoft and $\half\fboft$, are only valid for $\psq=0$, the 
 measurement of \ftqed is corrected for the effect of non-zero \psq in 
 the data by multiplying the result of the unfolding for $\pzm=0.05$~\gevsq 
 by the ratio of
 \ftqed for $\psq=0$ and \ftqed for $\pzm=0.05$~\gevsq as predicted by QED.
 Then \faqed and \fbqed are calculated by multiplying the measured ratios
 by \ftqed and corrected for the bin size effect as before.
 The corrected values for \ftqed, \faqed and \fbqed are listed in 
 Table~\ref{tab:aztot}, and \faqed and \fbqed are shown in 
 Figure~\ref{fig:pr271_15}. The QED predictions for \qsq = 5.4 \gevsq and 
 $\epsilon=\rho=1$ nicely describe the data.
%
%
\section{Summary and conclusions}
\label{sec:concl}
 The complete set of data collected by the OPAL experiment
 at centre-of-mass energies close to the mass of the \zn boson are used
 to extract information about the QED structure of the photon.
 The structure function \ftqed and the differential cross-section 
 \dsigdx for quasi-real photons are measured
 in the \qsq range from 1.5 to 400 \gevsq,
 the largest kinematic range ever covered by a single experiment. 
 The predictions of QED are found to be in good agreement with the data,
 and the predicted \psq suppression of the structure function \ftqed 
 is clearly observed.
 \par
 For the first time, the detailed quantitative analysis of the QED structure
 of the photon has been extended to highly virtual photons.
 Due to the non-vanishing interference terms \ttl and \ttt in the 
 kinematical region studied, strong cancellations in the differential 
 cross-section occur between the cross-section and interference terms.
 Consequently, no clear relation of a structure 
 function to the cross-section terms can be found.
 Therefore the differential cross-section for the reaction 
 $\ee\rightarrow\ee\mupmum$, proceeding via the exchange of two highly 
 virtual photons is measured instead, 
 for the range $1.5< \qsq < 30$ \gevsq and $1.5< \psq < 20$ \gevsq. 
 The QED predictions using the GALUGA program show good
 agreement with the data for the full cross-section
 and also the presence of the interference terms \ttl and \ttt in the data.
 \par
 The azimuthal correlations between electrons and muons are used to extract
 the structure functions \faqed and \fbqed in bins of $x$. 
 The results are consistent with those already published and with the
 QED predictions. 
 The measurements presented here supersede the earlier
 structure function results of OPAL~\cite{OPALPR182,OPALPR088}
%
%
\bigskip\bigskip\bigskip
\appendix
\par
Acknowledgements:
\par
 We wish to thank C.~Berger and G.~Schuler for valuable discussions
 concerning the interpretation of the measurements for highly virtual 
 photons and M.~Seymour for providing us with the structure function 
 calculations for the measurement of \faqed and \fbqed.
 \par
 We particularly wish to thank the SL Division for the efficient operation
 of the LEP accelerator at all energies
 and for their continuing close cooperation with
 our experimental group.  We thank our colleagues from CEA, DAPNIA/SPP,
 CE-Saclay for their efforts over the years on the time-of-flight and trigger
 systems which we continue to use.  In addition to the support staff at our own
 institutions we are pleased to acknowledge the  \\
 Department of Energy, USA, \\
 National Science Foundation, USA, \\
 Particle Physics and Astronomy Research Council, UK, \\
 Natural Sciences and Engineering Research Council, Canada, \\
 Israel Science Foundation, administered by the Israel
 Academy of Science and Humanities, \\
 Minerva Gesellschaft, \\
 Benoziyo Center for High Energy Physics,\\
 Japanese Ministry of Education, Science and Culture (the
 Monbusho) and a grant under the Monbusho International
 Science Research Program,\\
 Japanese Society for the Promotion of Science (JSPS),\\
 German Israeli Bi-national Science Foundation (GIF), \\
 Bundesministerium f\"ur Bildung, Wissenschaft,
 Forschung und Technologie, Germany, \\
 National Research Council of Canada, \\
 Research Corporation, USA,\\
 Hungarian Foundation for Scientific Research, OTKA T-016660, 
 T023793 and OTKA F-023259.\\
%
%

%
\pagebreak\bigskip
%
\renewcommand{\arraystretch}{1.1}
\begin{table}[htb]
\begin{center}\begin{tabular}{|c|c|c|c|c|}\cline{1-4}
 Sample        &      SW1 &      SW2 &\multicolumn{1}{||c|}{SW}       \\
 \cline{1-4}
 \qsq [\gevsq] &  1.5$-$3 &    3$-$7 &\multicolumn{1}{||c|}{1.5$-$7}  \\
 \psq [\gevsq] &      $-$ &      $-$ &\multicolumn{1}{||c|}{$-$}      \\ 
 $x$           & 0$-$0.97 & 0$-$0.97 &\multicolumn{1}{||c|}{0$-$0.97} \\
 $y$           & 0$-$0.5  & 0$-$0.5  &\multicolumn{1}{||c|}{0$-$0.5}   \\
 \cline{1-4}
 \qzm [\gevsq] &      2.2 &      4.2 &\multicolumn{1}{||c|}{3.0}      \\
 \pzm [\gevsq] &     0.05 &     0.05 &\multicolumn{1}{||c|}{0.05}     \\
 \cline{1-4}
 data       & 3259  $\pm$  57 & 2292 $\pm$  48 &
 \multicolumn{1}{||c|}{5551 $\pm$ 75} \\
 signal     & 3176  $\pm$  25 & 2273 $\pm$  21 & 
 \multicolumn{1}{||c|}{5449 $\pm$ 33} \\
 background & 47.9  $\pm$ 6.7 & 60.6 $\pm$ 9.2 & 
 \multicolumn{1}{||c|}{108.6 $\pm$ 11.4} \\
 \cline{1-4}
\multicolumn{5}{c}{} \\\cline{1-5}
 Sample        &      FD1 &      FD2 &      FD3 &
 \multicolumn{1}{||c|}{FD}\\\hline
 \qsq [\gevsq] &   6$-$10 &  10$-$15 &  15$-$30 &
 \multicolumn{1}{||c|}{6$-$30} \\
 \psq [\gevsq] &      $-$ &      $-$ &      $-$ &
 \multicolumn{1}{||c|}{$-$}\\
 $x$           & 0$-$0.97 & 0$-$0.97 & 0$-$0.97 &
 \multicolumn{1}{||c|}{0$-$0.97}\\
 $y$           & 0$-$0.5  & 0$-$0.5  & 0$-$0.5  &
 \multicolumn{1}{||c|}{0$-$0.5}\\ \hline
 \qzm [\gevsq] &      8.4 &     12.4 &     21.0 &
 \multicolumn{1}{||c|}{13.4} \\
 \pzm [\gevsq] &     0.05 &     0.05 &     0.05 &
 \multicolumn{1}{||c|}{0.05} \\\hline
 data      
 & 1058 $\pm$  33 &  790 $\pm$  28 &  719 $\pm$  27 &
 \multicolumn{1}{||c|}{2567 $\pm$  51}\\
 signal     
 &  988 $\pm$  16 &  803 $\pm$  14 &  738 $\pm$  13 &
 \multicolumn{1}{||c|}{2531 $\pm$  25} \\
 background 
 & 41.4 $\pm$ 4.0 & 39.5 $\pm$ 3.9 & 38.2 $\pm$ 3.9 &
 \multicolumn{1}{||c|}{119.1 $\pm$ 6.8}\\\hline
\multicolumn{5}{c}{} \\\hline
 Sample        &       EE &     DB1 &      DB2 &        DB3 \\
 \cline{1-5}
 \qsq [\gevsq] &  70$-$400 &  1.5$-$6 &   6$-$30 & 70$-$300 \\
 \psq [\gevsq] &       $-$ &  1.5$-$6 & 1.5$-$20 & 1.5$-$20 \\
 $x$           & 0.1$-$0.9 & 0$-$0.65 & 0$-$0.75 &    0$-$1 \\
 $y$           & 0$-$0.5   & 0$-$0.5  & 0$-$0.5  &   0$-$0.5 \\
 \cline{1-5}
 \qzm [\gevsq] &       130 &      3.6 &     14.0 &      140 \\
 \pzm [\gevsq] &      0.05 &      2.3 &      5.0 &       16 \\
\hline
 data   
 &   163 $\pm$ 12.8 &  111 $\pm$ 10.5 & 116   $\pm$ 10.8 & 8   $\pm$ 2.8 \\
 signal 
 & 161.9 $\pm$  4.7 & 85.0 $\pm$  2.5 & 102.1 $\pm$  2.7 & 7.9 $\pm$ 0.8 \\
 background
 &  17.4 $\pm$  2.7 &  6.4 $\pm$  3.1 &   7.5 $\pm$  1.4 & 2.0 $\pm$ 0.7 \\
 \cline{1-5}
 \end{tabular}
 \caption{The sample definitions in terms of \psq, \qsq, $x$ and $y$, the 
          average values \qzm and \pzm as obtained from the data 
          and Monte Carlo
          and the number of events selected from the data (data),
          together with the predictions of the signal Monte Carlo (signal) 
          and background events (background).
          The SW and the FD samples are not statistically 
          independent from the other samples, i.e. the 
          SW sample contains all events from SW1 and SW2 
          and the FD sample contains all events from FD1, FD2 and FD3
          }
 \label{tab:samp}
 \end{center}\end{table}
%
\renewcommand{\arraystretch}{1.1}
\begin{table}[htb]
\begin{center}\begin{tabular}{|c|cc||c|}\hline
     & SW1 & SW2  &SW  \\\cline{2-4}
 $x$ & \dsigdx [pb] & \dsigdx [pb]  &\dsigdx [pb]  \\\hline
 0.00$-$0.10
 & 409.0 $\pm$  23.6 $\pm$ 16.3
 & 164.4 $\pm$  15.0 $\pm$ 25.0
 & 562.3 $\pm$  28.3 $\pm$ 33.2
 \\
 0.10$-$0.20
 & 443.4 $\pm$  19.7 $\pm$ 16.7
 & 217.4 $\pm$  13.0 $\pm$ $\pz$8.2
 & 663.0 $\pm$  25.4 $\pm$ 22.3
 \\
 0.20$-$0.30
 & 423.2 $\pm$  18.8 $\pm$ 16.2
 & 223.8 $\pm$  12.3 $\pm$ $\pz$8.2
 & 647.9 $\pm$  24.4 $\pm$ 22.4
 \\
 0.30$-$0.40
 & 430.9 $\pm$  18.3 $\pm$ 13.8
 & 220.9 $\pm$  11.8 $\pm$ $\pz$6.6
 & 650.3 $\pm$  23.6 $\pm$ 21.4
 \\
 0.40$-$0.50
 & 377.7 $\pm$  17.6 $\pm$ 10.9
 & 185.7 $\pm$  10.1 $\pm$ $\pz$4.9
 & 556.3 $\pm$  21.6 $\pm$ 12.5
 \\
 0.50$-$0.60
 & 366.0 $\pm$  18.3 $\pm$ 10.3
 & 184.8 $\pm$  10.8 $\pm$ $\pz$5.5
 & 542.4 $\pm$  22.6 $\pm$ 15.0
 \\
 0.60$-$0.70
 & 380.1 $\pm$  19.9 $\pm$ 12.4
 & 198.3 $\pm$  11.1 $\pm$ $\pz$4.9
 & 578.7 $\pm$  24.3 $\pm$ 23.4
 \\
 0.70$-$0.80
 & 329.5 $\pm$  20.6 $\pm$ $\pz$7.8
 & 181.0 $\pm$  11.1 $\pm$ $\pz$3.6
 & 500.9 $\pm$  24.3 $\pm$ 12.3
 \\
 0.80$-$0.90
 & 302.5 $\pm$  19.4 $\pm$ $\pz$7.4
 & 166.7 $\pm$  11.1 $\pm$ $\pz$3.6
 & 467.7 $\pm$  23.9 $\pm$ 28.0
 \\
 0.90$-$0.97
 & 181.6 $\pm$  18.7 $\pm$ 41.9
 & 124.3 $\pm$  11.0 $\pm$ 13.5
 & 304.9 $\pm$  22.3 $\pm$ 45.5
 \\
 \hline
 \qzm  &   2.2 \gevsq &   4.2 \gevsq &   3.0 \gevsq \\
 \prob &  0.79        &  0.26        &  0.46 \\
 \hline
 \end{tabular}
 \caption{The measured differential cross-section \dsigdx for the SW samples.
          In addition, the average \qsq value for each sample, \qzm, and the 
          $\chi^2$ probabilities, \prob, are listed.
          The sample definitions are given in Table\protect~\ref{tab:samp}.
          The SW sample contains all events from SW1 and SW2. 
          The first error is statistical and the second systematic
         }
 \label{tab:sf2sw}
 \end{center}\end{table}
%
\renewcommand{\arraystretch}{1.1}
\begin{table}[htb]
\begin{center}\begin{tabular}{|c|c||c|}\hline
     & FD1 & FD   \\\cline{2-3}
 $x$ &\dsigdx [pb]  &\dsigdx [pb] \\\hline
 0.00$-$0.10
 &      35.4 $\pm$  4.8 $\pm$ 2.8
 & $\pz$61.2 $\pm$  6.7 $\pm$ 4.2
 \\
 0.10$-$0.20
 &  69.4 $\pm$  5.6 $\pm$ 4.9
 & 116.2 $\pm$  8.1 $\pm$ 6.0
 \\
 0.20$-$0.30
 &  66.2 $\pm$  5.8 $\pm$ 4.0
 & 109.6 $\pm$  7.9 $\pm$ 5.0
 \\
 0.30$-$0.40
 &  67.9 $\pm$  5.5 $\pm$ 3.6
 & 123.2 $\pm$  8.1 $\pm$ 5.7
 \\
 0.40$-$0.50
 &  71.2 $\pm$  5.5 $\pm$ 3.8
 & 111.5 $\pm$  7.5 $\pm$ 4.4
 \\
 0.50$-$0.60
 &  71.3 $\pm$  5.4 $\pm$ 3.3
 & 124.2 $\pm$  7.5 $\pm$ 5.1
 \\
 0.60$-$0.70
 &  65.3 $\pm$  5.4 $\pm$ 2.5
 & 134.5 $\pm$  7.7 $\pm$ 5.7
 \\
 0.70$-$0.80
 &  64.8 $\pm$  5.3 $\pm$ 2.2
 & 128.3 $\pm$  7.4 $\pm$ 4.3
 \\
 0.80$-$0.90
 &  72.1 $\pm$  5.8 $\pm$ 3.5
 & 124.2 $\pm$  7.6 $\pm$ 4.8
 \\
 0.90$-$0.97
 &  57.5 $\pm$  5.6 $\pm$ 3.7
 & 104.6 $\pm$  7.0 $\pm$ 5.8
 \\
 \hline
 \qzm  &   8.4 \gevsq & 13.4 \gevsq \\
 \prob &  0.61        &  0.13 \\
 \hline
 \multicolumn{3}{c}{} \\
 \end{tabular}
 \begin{tabular}{|c|cc|}
 \hline
     & FD2 & FD3  \\\cline{2-3}
 $x$ &\dsigdx [pb]  &\dsigdx [pb] \\\hline
 0.00$-$0.15
 &  20.2 $\pm$  2.9 $\pm$ 1.6
 &  11.5 $\pm$  2.7 $\pm$ 1.2
 \\
 0.15$-$0.30
 &  26.4 $\pm$  3.0 $\pm$ 1.5
 &  21.9 $\pm$  2.8 $\pm$ 1.5
 \\
 0.30$-$0.45
 &  27.6 $\pm$  2.8 $\pm$ 1.7
 &  22.9 $\pm$  2.9 $\pm$ 1.6
 \\
 0.45$-$0.60
 &  24.7 $\pm$  2.5 $\pm$ 1.9
 &  26.9 $\pm$  2.8 $\pm$ 1.4
 \\
 0.60$-$0.75
 &  35.7 $\pm$  2.9 $\pm$ 1.9
 &  30.5 $\pm$  2.7 $\pm$ 1.3
 \\
 0.75$-$0.90
 &  27.9 $\pm$  2.8 $\pm$ 1.1
 &  28.9 $\pm$  2.5 $\pm$ 1.1
 \\
 0.90$-$0.97
 &  24.5 $\pm$  2.9 $\pm$ 1.0
 &  23.6 $\pm$  3.0 $\pm$ 1.0
 \\
 \hline
 \qzm  &  12.4 \gevsq & 21.0 \gevsq \\
 \prob &  0.07        &  0.09 \\
 \hline
 \end{tabular}
 \caption{The differential cross-section \dsigdx for the FD samples.
          In addition, the average \qsq value for each sample, \qzm, and the 
          $\chi^2$ probabilities, \prob, are listed.
          The sample definitions are given in Table\protect~\ref{tab:samp}.
          The FD sample contains all events from FD1, FD2 and FD3.
          The first error is statistical and the second systematic
         }
 \label{tab:sf2fd}
 \end{center}\end{table}
%
\renewcommand{\arraystretch}{1.1}
\begin{table}[htb]
\begin{center}\begin{tabular}{|c|c|}\hline
     & EE \\\cline{2-2}
 $x$ &\dsigdx [pb] \\\hline
 0.1$-$0.4 &  2.78 $\pm$  0.76 $\pm$ 0.28 \\
 0.4$-$0.6 &  4.27 $\pm$  0.58 $\pm$ 0.38 \\
 0.6$-$0.8 &  5.79 $\pm$  0.67 $\pm$ 0.39 \\
 0.8$-$0.9 &  5.88 $\pm$  0.68 $\pm$ 0.30 \\
 \hline
 \qzm  &   130 \gevsq \\
 \prob &  0.48 \\
 \hline
 \end{tabular}
 \caption{The measured differential cross-section \dsigdx for the EE sample.
          In addition, the average \qsq value for the sample, \qzm, and the 
          $\chi^2$ probability, \prob, are listed.
          The sample definition is given in Table\protect~\ref{tab:samp}.
          The first error is statistical and the second systematic
         }
 \label{tab:sf2ee}
 \end{center}\end{table}
%
\renewcommand{\arraystretch}{1.1}
\begin{table}[htb]
\begin{center}\begin{tabular}{|c|c|c|c|}\hline
     & DB1          &     & DB2         \\\cline{2-2}\cline{4-4}
 $x$ & \dsigdx [pb] & $x$ & \dsigdx [pb] \\\hline
 0.00$-$0.20 &  $\pz$9.77 $\pm$  1.62 $\pm$ 0.78 &
 0.00$-$0.25 &       5.26 $\pm$  0.82 $\pm$ 0.99 \\
 0.20$-$0.40 &      10.45 $\pm$  1.26 $\pm$ 0.59  &
 0.25$-$0.50 &       6.87 $\pm$  0.78 $\pm$ 0.74 \\
 0.40$-$0.65 &  $\pz$4.34 $\pm$  1.07 $\pm$ 0.21 &
 0.50$-$0.75 &       2.75 $\pm$  0.60 $\pm$ 0.22 \\
 \hline
 \qzm  &   3.6 \gevsq & \qzm  & 14.0 \gevsq \\
 \pzm  &   2.3 \gevsq & \pzm  &  5.0 \gevsq \\
 \prob &  0.26        & \prob &  0.32 \\
 \hline
 \end{tabular}
 \caption{The measured differential cross-section \dsigdx for the DB samples.
          In addition, the average \qsq and \psq values for each sample, 
          \qzm and \pzm, and the $\chi^2$ probabilities, \prob, are listed.
          The sample definitions are given in Table\protect~\ref{tab:samp}.
          The first error is statistical and the second systematic
         }
 \label{tab:sf2db}
 \end{center}\end{table}
%
\renewcommand{\arraystretch}{1.1}
\begin{table}[htb]
\begin{center}\begin{tabular}{|c|cc||c|}\hline
      & SW1 & SW2  &SW  \\\cline{2-4}
 $x$  & \ftqed & \ftqed &\ftqed \\\hline
 0.00$-$0.10
 & 0.115 $\pm$  0.007 $\pm$ 0.005
 & 0.108 $\pm$  0.010 $\pm$ 0.016
 & 0.113 $\pm$  0.006 $\pm$ 0.007
 \\
 0.10$-$0.20
 & 0.219 $\pm$  0.010 $\pm$ 0.008
 & 0.237 $\pm$  0.014 $\pm$ 0.009
 & 0.230 $\pm$  0.009 $\pm$ 0.008
 \\
 0.20$-$0.30
 & 0.282 $\pm$  0.012 $\pm$ 0.011
 & 0.320 $\pm$  0.018 $\pm$ 0.012
 & 0.300 $\pm$  0.011 $\pm$ 0.010
 \\
 0.30$-$0.40 
 & 0.347 $\pm$  0.015 $\pm$ 0.011
 & 0.378 $\pm$  0.020 $\pm$ 0.011
 & 0.363 $\pm$  0.013 $\pm$ 0.012
 \\
 0.40$-$0.50
 & 0.356 $\pm$  0.017 $\pm$ 0.010
 & 0.373 $\pm$  0.020 $\pm$ 0.010
 & 0.364 $\pm$  0.014 $\pm$ 0.008
 \\
 0.50$-$0.60
 & 0.400 $\pm$  0.020 $\pm$ 0.011
 & 0.421 $\pm$  0.025 $\pm$ 0.012
 & 0.409 $\pm$  0.017 $\pm$ 0.011
 \\
 0.60$-$0.70
 & 0.483 $\pm$  0.025 $\pm$ 0.016
 & 0.519 $\pm$  0.029 $\pm$ 0.013
 & 0.507 $\pm$  0.021 $\pm$ 0.020
 \\
 0.70$-$0.80
 & 0.491 $\pm$  0.031 $\pm$ 0.012
 & 0.556 $\pm$  0.034 $\pm$ 0.011
 & 0.516 $\pm$  0.025 $\pm$ 0.013
 \\
 0.80$-$0.90
 & 0.532 $\pm$  0.034 $\pm$ 0.013
 & 0.601 $\pm$  0.040 $\pm$ 0.013
 & 0.574 $\pm$  0.029 $\pm$ 0.034
 \\
 0.90$-$0.97
 & 0.308 $\pm$  0.032 $\pm$ 0.071
 & 0.470 $\pm$  0.041 $\pm$ 0.051
 & 0.397 $\pm$  0.029 $\pm$ 0.059
 \\
 \hline
 \qzm  &   2.2 \gevsq &   4.2 \gevsq &   3.0 \gevsq \\
 \prob &  0.79        &  0.26        &  0.46 \\
 \hline
 \end{tabular}
 \caption{The measured structure function \ftqed for the SW samples.
          In addition, the average \qsq value for each sample, \qzm, and the 
          $\chi^2$ probabilities, \prob, are listed.
          The sample definitions are given in Table\protect~\ref{tab:samp}.
          The SW sample contains all events from SW1 and SW2. 
          The first error is statistical and the second systematic
         }
 \label{tab:ff2sw}
 \end{center}\end{table}
%
\renewcommand{\arraystretch}{1.1}
\begin{table}[htb]
\begin{center}\begin{tabular}{|c|c||c|}\hline
     & FD1    & FD     \\\cline{2-3}
 $x$ & \ftqed & \ftqed \\\hline
 0.00$-$0.10
 & 0.090 $\pm$  0.012 $\pm$ 0.007
 & 0.095 $\pm$  0.010 $\pm$ 0.007 
 \\
 0.10$-$0.20
 & 0.271 $\pm$  0.022 $\pm$ 0.019
 & 0.264 $\pm$  0.018 $\pm$ 0.014 
 \\
 0.20$-$0.30
 & 0.334 $\pm$  0.029 $\pm$ 0.020
 & 0.319 $\pm$  0.023 $\pm$ 0.014 
 \\
 0.30$-$0.40
 & 0.409 $\pm$  0.033 $\pm$ 0.022
 & 0.428 $\pm$  0.028 $\pm$ 0.020 
 \\
 0.40$-$0.50
 & 0.496 $\pm$  0.038 $\pm$ 0.026
 & 0.446 $\pm$  0.030 $\pm$ 0.018 
 \\
 0.50$-$0.60
 & 0.563 $\pm$  0.043 $\pm$ 0.026
 & 0.558 $\pm$  0.034 $\pm$ 0.023 
 \\
 0.60$-$0.70
 & 0.596 $\pm$  0.049 $\pm$ 0.023
 & 0.698 $\pm$  0.040 $\pm$ 0.030 
 \\
 0.70$-$0.80
 & 0.687 $\pm$  0.056 $\pm$ 0.023
 & 0.770 $\pm$  0.044 $\pm$ 0.026 
 \\
 0.80$-$0.90
 & 0.891 $\pm$  0.072 $\pm$ 0.044
 & 0.871 $\pm$  0.053 $\pm$ 0.033 
 \\
 0.90$-$0.97
 & 0.761 $\pm$  0.074 $\pm$ 0.049
 & 0.795 $\pm$  0.053 $\pm$ 0.044 
 \\
 \hline
 \qzm  &   8.4 \gevsq & 13.4 \gevsq \\
 \prob &  0.61        &  0.13 \\
 \hline
 \multicolumn{3}{c}{} \\\cline{1-3}
 \end{tabular}
 \begin{tabular}{|c|cc|}
     & FD2    & FD3     \\\cline{2-3}
 $x$ & \ftqed & \ftqed \\\hline
 0.00$-$0.15
 & 0.151 $\pm$  0.022 $\pm$ 0.012
 & 0.117 $\pm$  0.028 $\pm$ 0.012
 \\
 0.15$-$0.30
 & 0.297 $\pm$  0.033 $\pm$ 0.017
 & 0.302 $\pm$  0.039 $\pm$ 0.021
 \\
 0.30$-$0.45
 & 0.402 $\pm$  0.041 $\pm$ 0.025
 & 0.403 $\pm$  0.051 $\pm$ 0.029
 \\
 0.45$-$0.60
 & 0.434 $\pm$  0.044 $\pm$ 0.034
 & 0.559 $\pm$  0.058 $\pm$ 0.030
 \\
 0.60$-$0.75
 & 0.758 $\pm$  0.062 $\pm$ 0.041
 & 0.782 $\pm$  0.070 $\pm$ 0.034
 \\
 0.75$-$0.90
 & 0.723 $\pm$  0.072 $\pm$ 0.028
 & 0.907 $\pm$  0.080 $\pm$ 0.033
 \\
 0.90$-$0.97
 & 0.714 $\pm$  0.085 $\pm$ 0.030
 & 0.802 $\pm$  0.103 $\pm$ 0.033
 \\
 \hline
 \qzm  &  12.4 \gevsq & 21.0 \gevsq \\
 \prob &  0.07        &  0.09 \\
 \hline
 \end{tabular}
 \caption{The measured structure function \ftqed for the FD samples.
          In addition, the average \qsq value for each sample, \qzm, and the 
          $\chi^2$ probabilities, \prob, are listed.
          The sample definitions are given in Table\protect~\ref{tab:samp}.
          The FD sample contains all events from FD1, FD2 and FD3.
          The first error is statistical and the second systematic
         }
 \label{tab:ff2fd}
 \end{center}\end{table}
%
\renewcommand{\arraystretch}{1.1}
\begin{table}[htb]
\begin{center}\begin{tabular}{|c|c|}\hline
     & EE \\\cline{2-2}
 $x$ & \ftqed  \\\hline
 0.1$-$0.4 & 0.343 $\pm$  0.094 $\pm$ 0.034 \\
 0.4$-$0.6 & 0.578 $\pm$  0.079 $\pm$ 0.052 \\
 0.6$-$0.8 & 0.936 $\pm$  0.109 $\pm$ 0.063 \\
 0.8$-$0.9 & 1.125 $\pm$  0.130 $\pm$ 0.057 \\
 \hline
 \qzm  &   130 \gevsq \\
 \prob &  0.48 \\
 \hline
 \end{tabular}
 \caption{The measured structure function \ftqed for the EE sample.
          In addition, the average \qsq value for the sample, \qzm, and the 
          $\chi^2$ probability, \prob, are listed.
          The sample definition is given in Table\protect~\ref{tab:samp}.
          The first error is statistical and the second systematic
         }
 \label{tab:ff2ee}
 \end{center}\end{table}
%
\renewcommand{\arraystretch}{1.1}
\begin{table}[htb]
\begin{center}\begin{tabular}{|c|c|c|}\hline
           $x$&                       \faoft  &            $\half\fboft$\\
\hline
      $x<0.25$& 
               \pz 0.176 $\pm$ 0.031 $\pm$ 0.010  &
                   0.075 $\pm$ 0.025 $\pm$ 0.008  \\
 0.25 $-$ 0.50& 
               \pz 0.018 $\pm$ 0.028 $\pm$ 0.008  &
                   0.099 $\pm$ 0.024 $\pm$ 0.010  \\
 0.50 $-$ 0.75& 
                $-$0.171 $\pm$ 0.029 $\pm$ 0.007  &
                   0.081 $\pm$ 0.027 $\pm$ 0.011  \\
      $x>0.75$& 
                $-$0.228 $\pm$ 0.037 $\pm$ 0.014  &
                   0.037 $\pm$ 0.033 $\pm$ 0.011  \\
\hline
 \end{tabular}
 \caption{The measured values of the structure function ratios 
          \faoft and $\half\fboft$ for the combined SW and FD sample. 
          The sample definitions are given in Table\protect~\ref{tab:samp}.
          The first error is statistical and the second systematic
         }
 \label{tab:azre}
 \end{center}\end{table}
%
\renewcommand{\arraystretch}{1.1}
\begin{table}[htb]
\begin{center}\begin{tabular}{|c|c|c|c|}\hline
           $x$&  \ftqed  &   \faqed  &     \fbqed\\
\hline
     $x<0.25$ &    0.249 $\pm$ 0.006 $\pm$ 0.008  &
               \pz 0.039 $\pm$ 0.007 $\pm$ 0.003  &
                   0.029 $\pm$ 0.010 $\pm$ 0.003  \\
 0.25 $-$ 0.50&    0.523 $\pm$ 0.011 $\pm$ 0.014  &
               \pz 0.011 $\pm$ 0.016 $\pm$ 0.004  &
                   0.101 $\pm$ 0.025 $\pm$ 0.011  \\
 0.50 $-$ 0.75&    0.738 $\pm$ 0.017 $\pm$ 0.019  &
                $-$0.122 $\pm$ 0.021 $\pm$ 0.006  &
                   0.121 $\pm$ 0.041 $\pm$ 0.017  \\
     $x>0.75$ &    0.871 $\pm$ 0.027 $\pm$ 0.021  &
                $-$0.201 $\pm$ 0.033 $\pm$ 0.013  &
                   0.063 $\pm$ 0.056 $\pm$ 0.018  \\
\hline
 \end{tabular}
 \caption{The measured values of the structure functions
          \ftqed, \faqed and \fbqed for the combined SW and FD sample. 
          The sample definitions are given in Table\protect~\ref{tab:samp}.
          The first error is statistical and the second systematic
         }
 \label{tab:aztot}
 \end{center}\end{table}
%
\clearpage
\pagebreak\bigskip
%
%
\begin{figure}
\begin{center}
\mbox{\epsfig{file=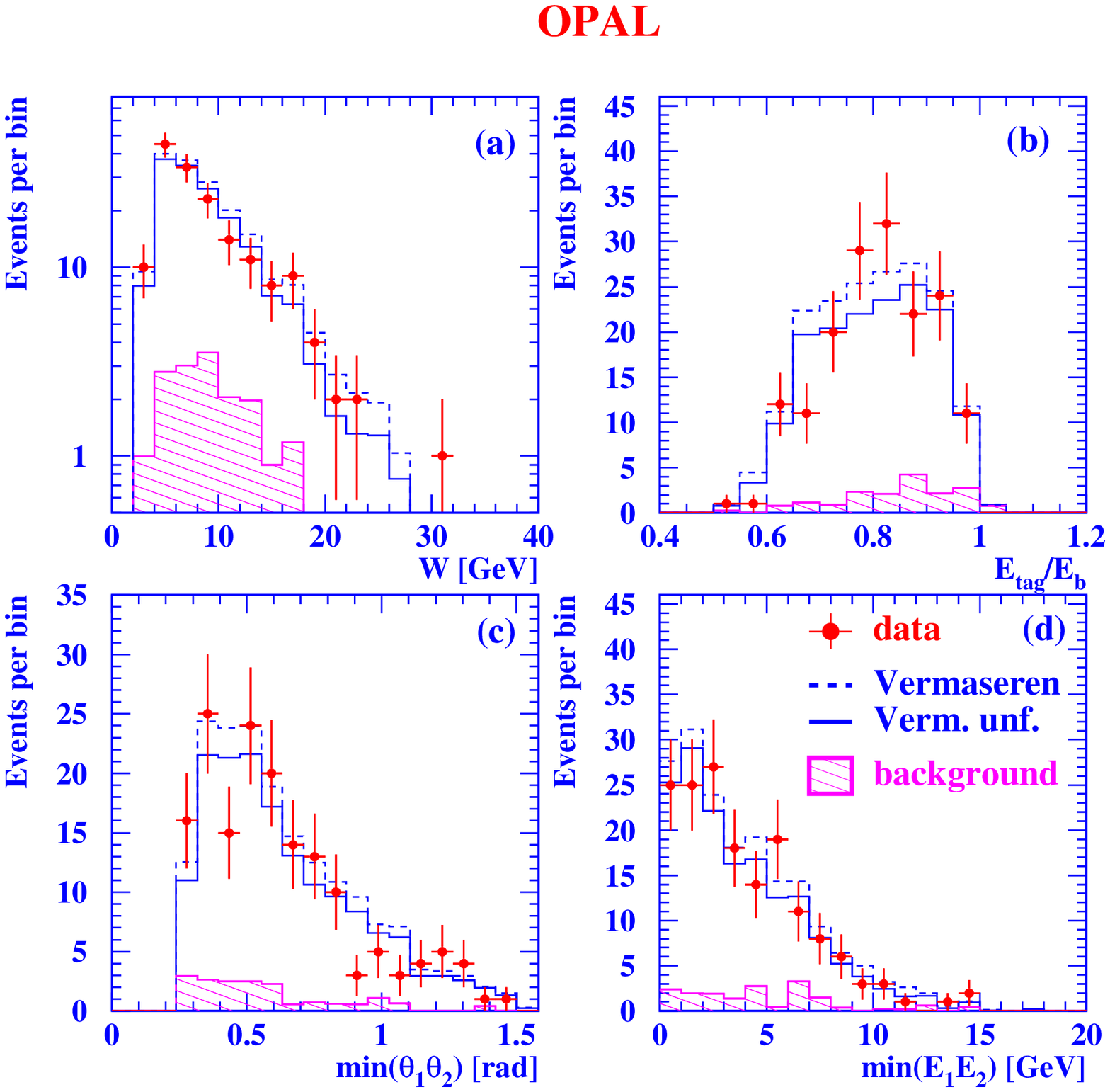,height=17cm}}
\caption{\label{fig:pr271_03}
 Comparison of the data with the Monte Carlo predictions for the EE sample.
 Shown are 
 (a) the measured invariant mass, $W$, 
 (b) the energy of the detected electron normalised to the energy of the 
 beam electrons, \etag/\eb,
 (c) the polar angle of the muon which is scattered closer to the beam 
 direction, \tmin,
 and
 (d) the energy of the muon with lower energy, \emin.
 The points represent the data with the statistical error only.
 The dashed lines denote the distributions of the Vermaseren 
 Monte Carlo with the background added to it before the unfolding 
 is performed, the full lines are the same distributions after 
 reweighting based on the unfolding procedure, 
 and the hatched histograms represent the background contribution
 }
\end{center}
\end{figure}
%
%
\begin{figure}
\begin{center}
\mbox{\epsfig{file=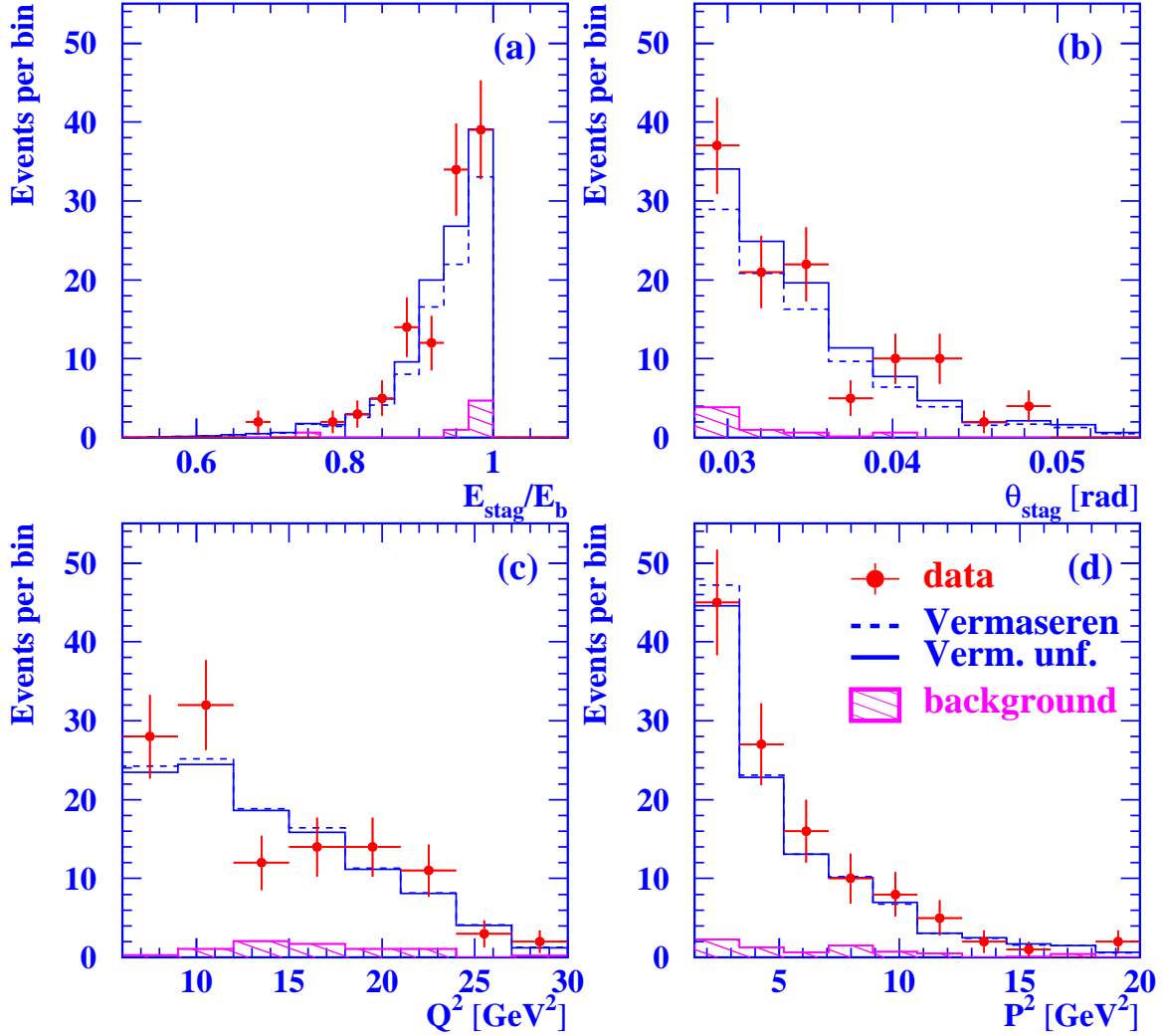,height=17cm}}
\caption{\label{fig:pr271_04}
 Comparison of the data with the Monte Carlo predictions 
 for the DB samples.
 Figures (a,b) and (c,d) are for the DB1 and DB2 samples, respectively.
 Shown are 
 (a) the energy of the second detected electron normalised 
 to the energy of the beam electrons, \estag/\eb, 
 (b) the polar angle of the second detected electron, \tstag,
 (c) the measured \qsq,
 and 
 (d) the measured \psq.
 The points represent the data with the statistical error only.
 The dashed lines denote the distributions of the Vermaseren 
 Monte Carlo with the background added to it before the unfolding 
 is performed, the full lines are the same distributions after 
 reweighting based on the unfolding procedure,
 and the hatched histograms represent the background contribution
 }
\end{center}
\end{figure}
%
%
\begin{figure}
\begin{center}
\mbox{\epsfig{file=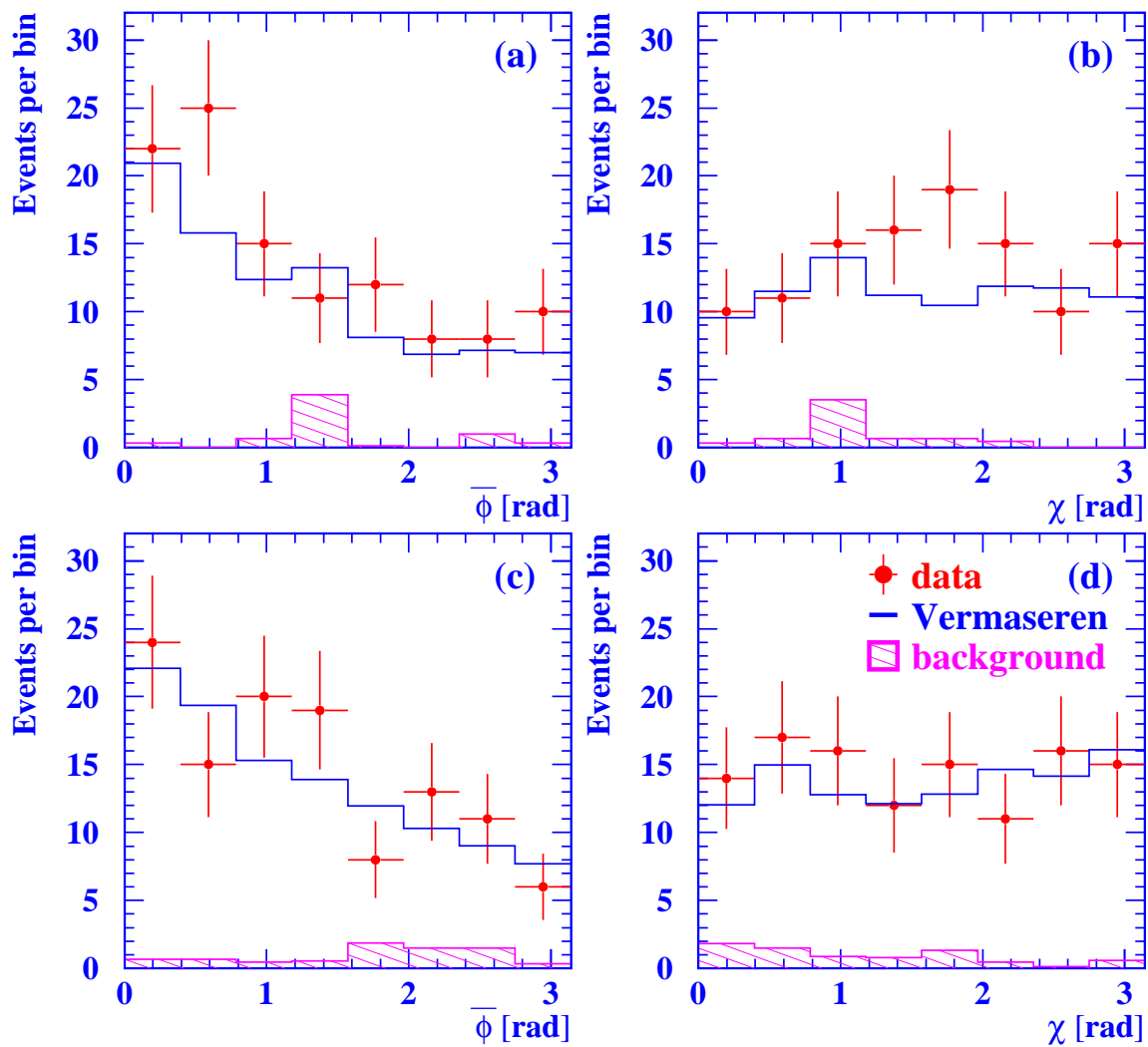,height=17cm}}
\caption{\label{fig:pr271_05}
 Comparison of the data with the Monte Carlo predictions 
 for the DB samples.
 Figures (a,b) and (c,d) are for the DB1 and DB2 samples, respectively.
 Shown are 
 (a,c) the angle \barph between the two scattering planes 
 of the electrons in the photon-photon centre-of-mass system,
 and
 (b,d) the measured azimuthal angle, $\chi$, defined with respect to the 
 plane containing the electron which radiated the photon of higher virtuality.
 The points represent the data with the statistical error only.
 The full lines denote the distributions of the Vermaseren Monte Carlo
 with the background added to it,
 and the hatched histograms represent the background contribution
 }
\end{center}
\end{figure}
%
%
\begin{figure}
\begin{center}
\mbox{\epsfig{file=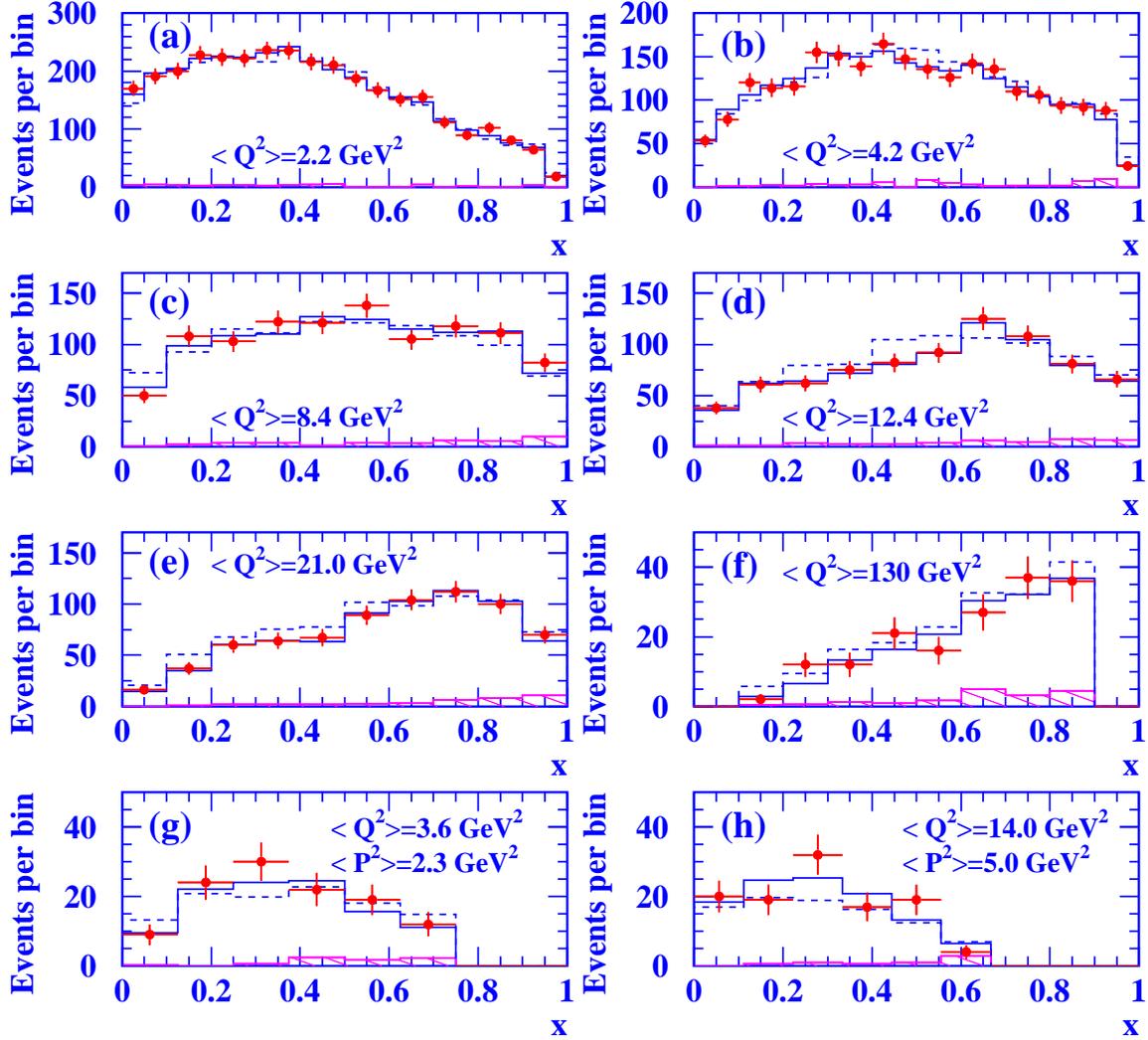,height=17cm}}
\caption{\label{fig:pr271_06}
 The measured $x$ distributions for the various samples
 (a) SW1, (b) SW2, (c) FD1, (d) FD2, (e) FD3, (f) EE, (g) DB1 and (h) DB2.
 The sample definitions are given in Table\protect~\ref{tab:samp}.
 The points represent the data with the statistical error only.
 The dashed lines denote the distributions of the Vermaseren
 Monte Carlo with the background added to it before the unfolding 
 is performed, the full lines are the same distributions after 
 reweighting based on the unfolding procedure, 
 and the hatched histograms represent the background contribution
  }
\end{center}
\end{figure}
%
%
\begin{figure}
\begin{center}
\mbox{\epsfig{file=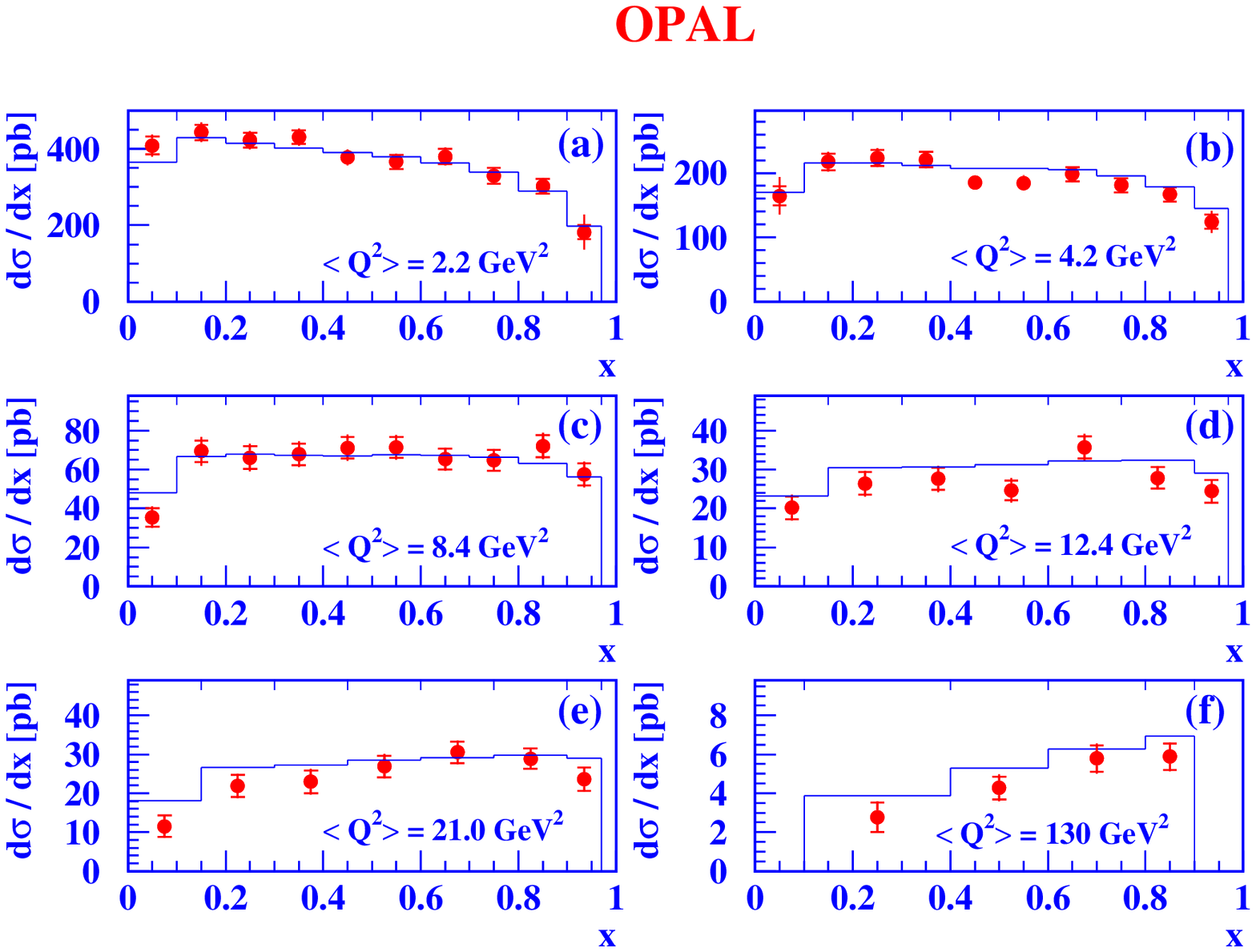,height=12.0cm}}
\caption{\label{fig:pr271_07}
  Differential cross-sections \dsigdx unfolded from the data for 
  the various independent samples of singly-tagged events.
  The samples shown are
  (a) SW1, (b) SW2, (c) FD1, (d) FD2, (e) FD3 and (f) EE.
  The points represent the data with their statistical (inner error bars) 
  and total errors (outer error bars). 
  The full line is the differential cross-sections as predicted by the 
  Vermaseren Monte Carlo.
  The tic marks at the top of the figures indicate the bin boundaries
  }
\end{center}
\end{figure}
%
%
\begin{figure}
\begin{center}
\mbox{\epsfig{file=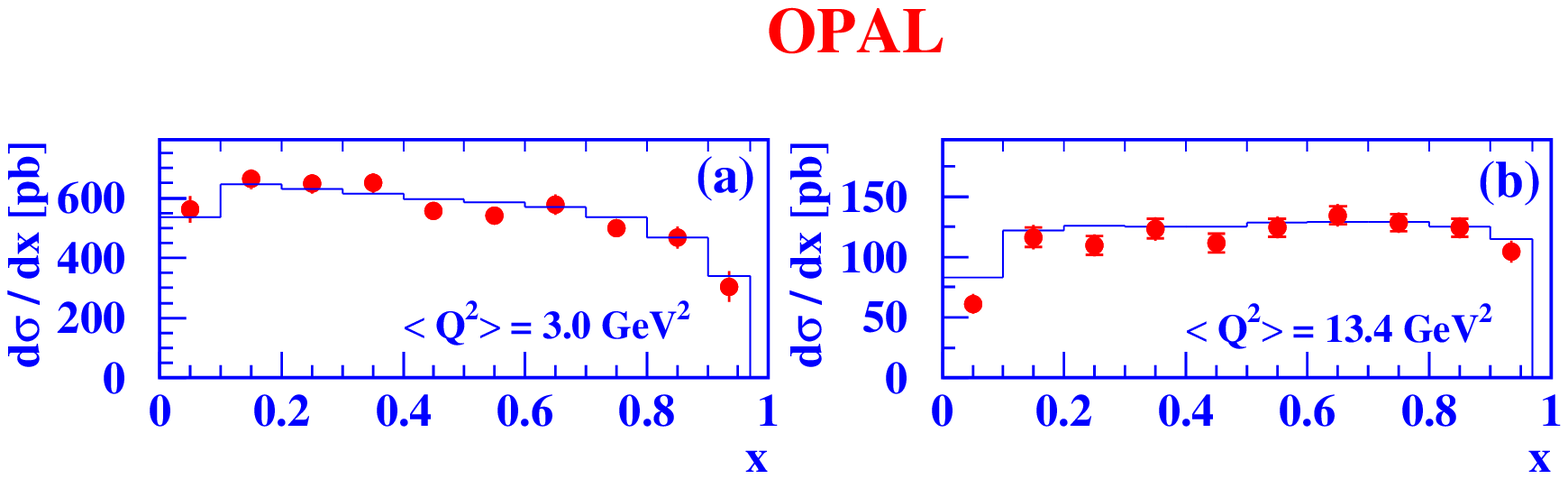,height=4.2cm}}
\caption{\label{fig:pr271_08}
  Differential cross-sections \dsigdx unfolded from the data for 
  the two combined samples of singly-tagged events.
  The samples shown are (a) SW and (b) FD.
  The symbols are as defined in Figure\protect~\ref{fig:pr271_07}
  }
\end{center}
\end{figure}
%
%
\begin{figure}
\begin{center}
\mbox{\epsfig{file=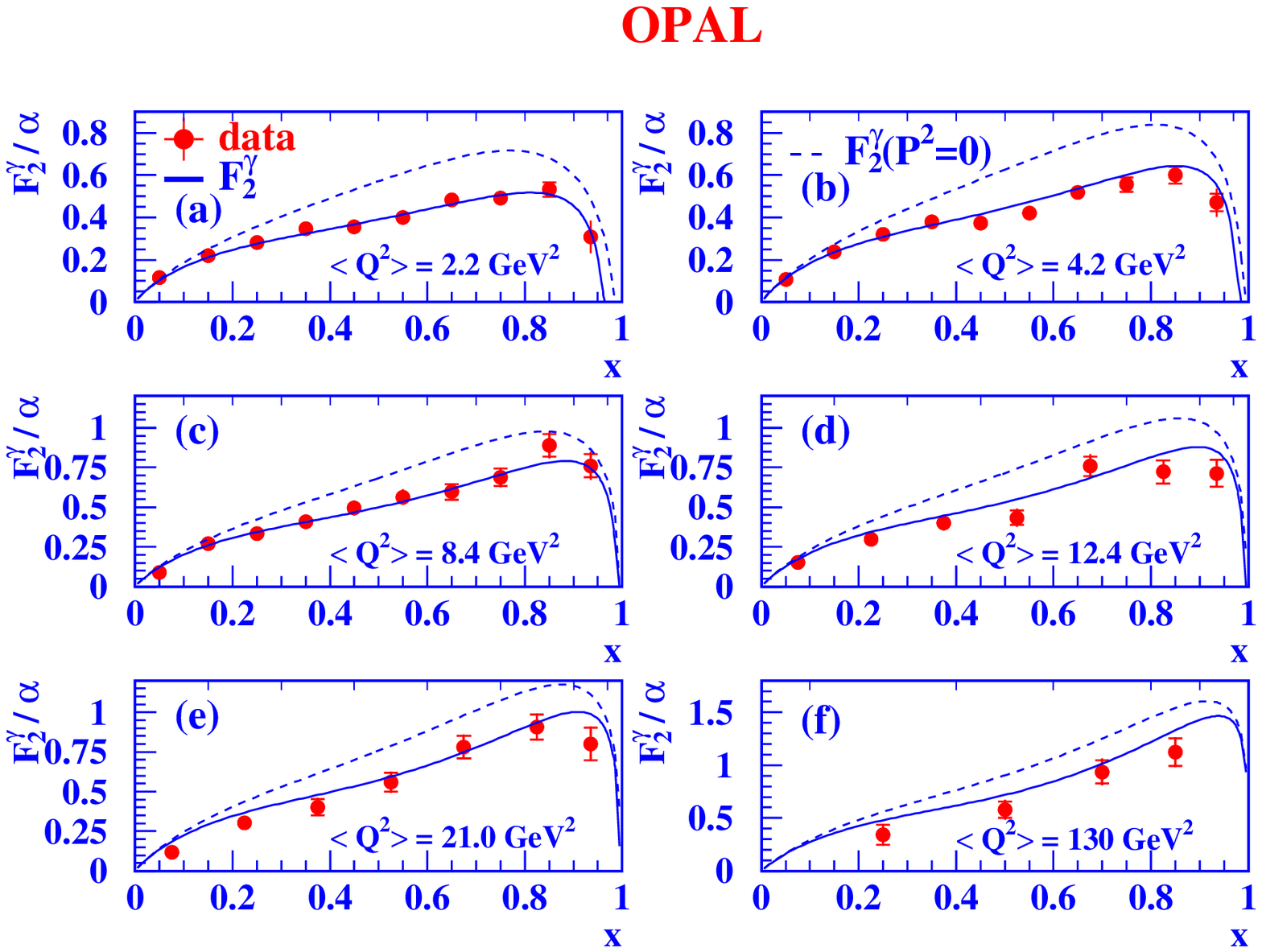,height=11cm}}
\caption{\label{fig:pr271_09}
  Structure functions \ftqed normalised by the fine structure constant
  \aem for the various independent samples of singly-tagged events.
  The ordering of 
  sub-figures is the same as in Figure\protect~\ref{fig:pr271_07}.
  The points represent the data with their statistical (inner error bars) 
  and total errors (outer error bars). 
  The full line is the structure function \ftqed normalised by the fine 
  structure constant \aem for the \qzm and \pzm values listed 
  in Table\protect~\ref{tab:samp}, and the dashed line shows the 
  structure function \ftqed normalised by the fine structure constant
  \aem for the same \qzm but for $\psq = 0$.
  The tic marks at the top of the figures indicate the bin boundaries
  }
\end{center}
\end{figure}
%
%
\begin{figure}
\begin{center}
\mbox{\epsfig{file=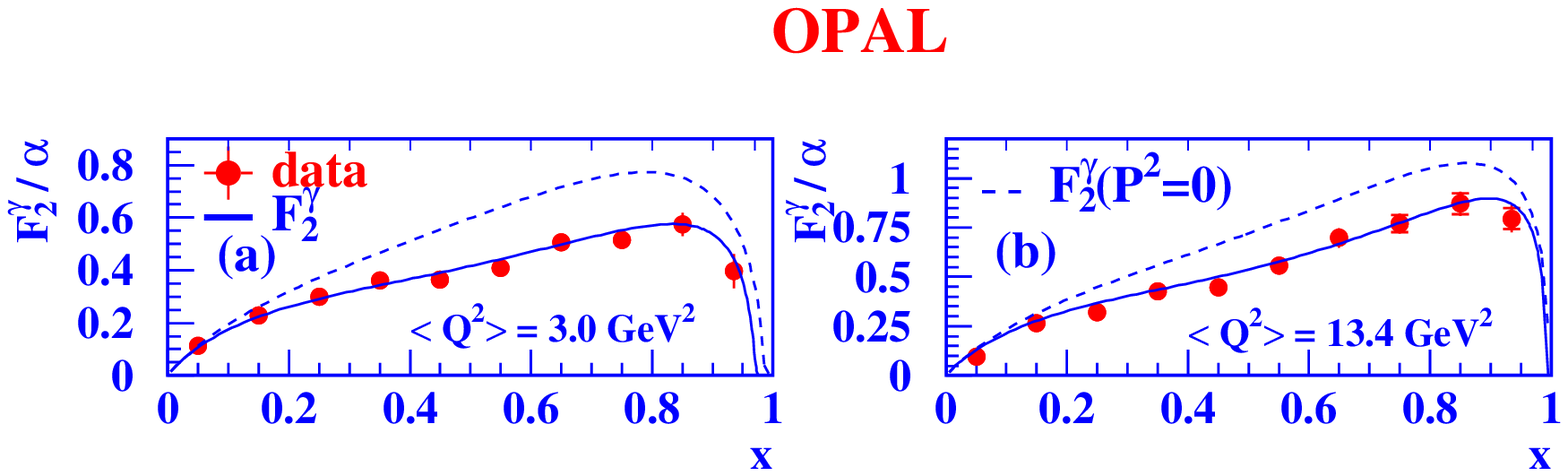,height=4.2cm}}
\caption{\label{fig:pr271_10}
  Structure functions \ftqed normalised by the fine structure constant
  \aem for the two combined samples of singly-tagged events.
  The samples shown are (a) SW and (b) FD.
  The symbols are as defined in Figure\protect~\ref{fig:pr271_09}
  }
\end{center}
\end{figure}
%
%
\begin{figure}
\begin{center}
\mbox{\epsfig{file=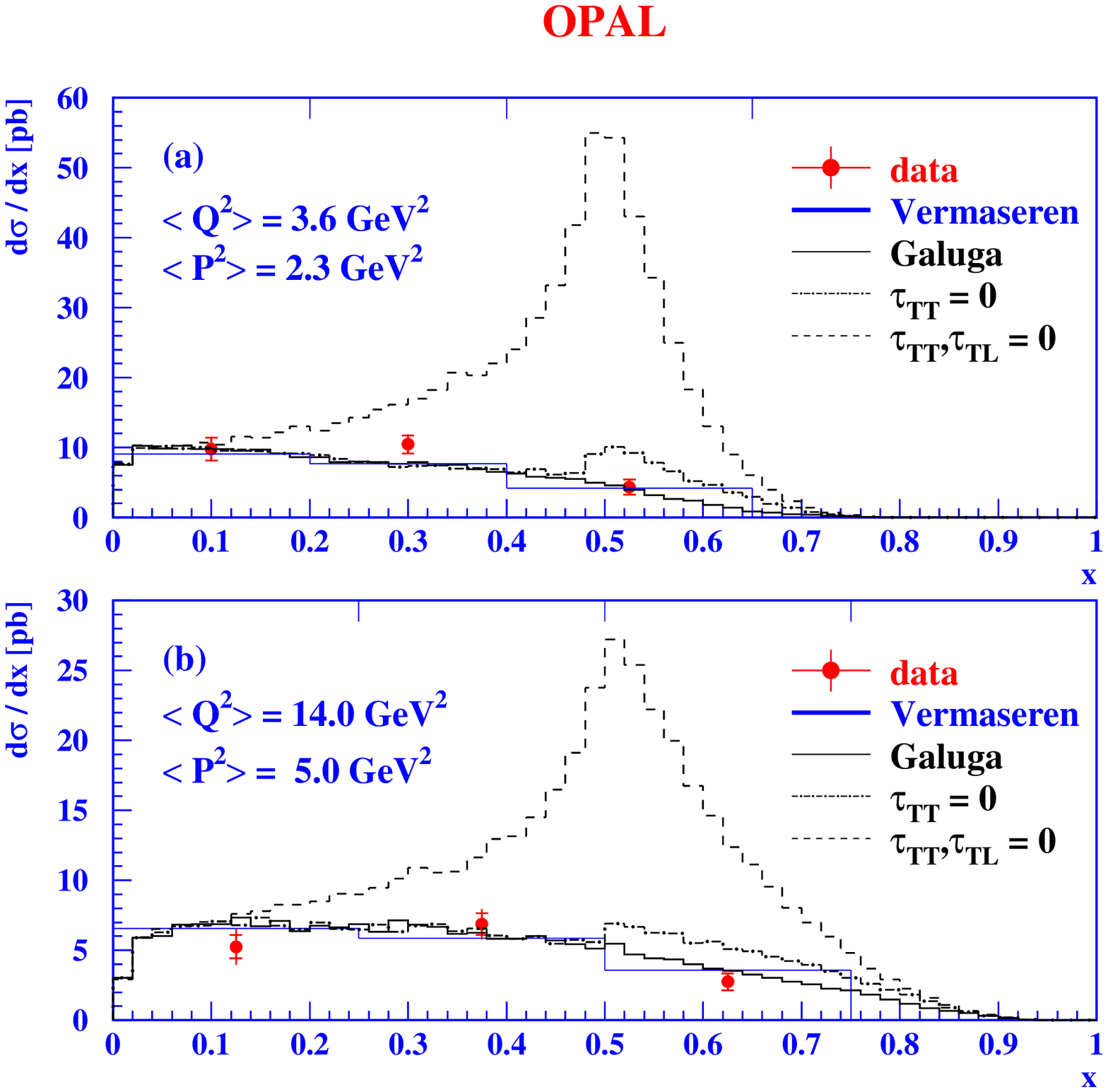,height=17cm}}
\caption{\label{fig:pr271_11}
  Differential cross-sections \dsigdx
  unfolded from the data, (a) for the DB1 sample and  (b) for the DB2 sample.
  The sample definitions are given in Table\protect~\ref{tab:samp}.
  The points represent the data with their statistical (inner error bars) 
  and total errors (outer error bars). 
  The full line denotes the differential cross-sections as predicted by the 
  Vermaseren Monte Carlo using the same bins as for the data.
  The additional three histograms represent the cross-section calculations
  from the GALUGA Monte Carlo for three different scenarios:
  the full cross-section (full line), the cross-section obtained for 
  vanishing \ttt (dot-dash) and the cross-section obtained for 
  vanishing \ttt and \ttl (dash), as defined in Eq.\protect~\ref{eqn:true}.
  The tic marks at the top of the figures indicate the bin boundaries
  }
\end{center}
\end{figure}
%
%
\begin{figure}
\begin{center}
\mbox{\epsfig{file=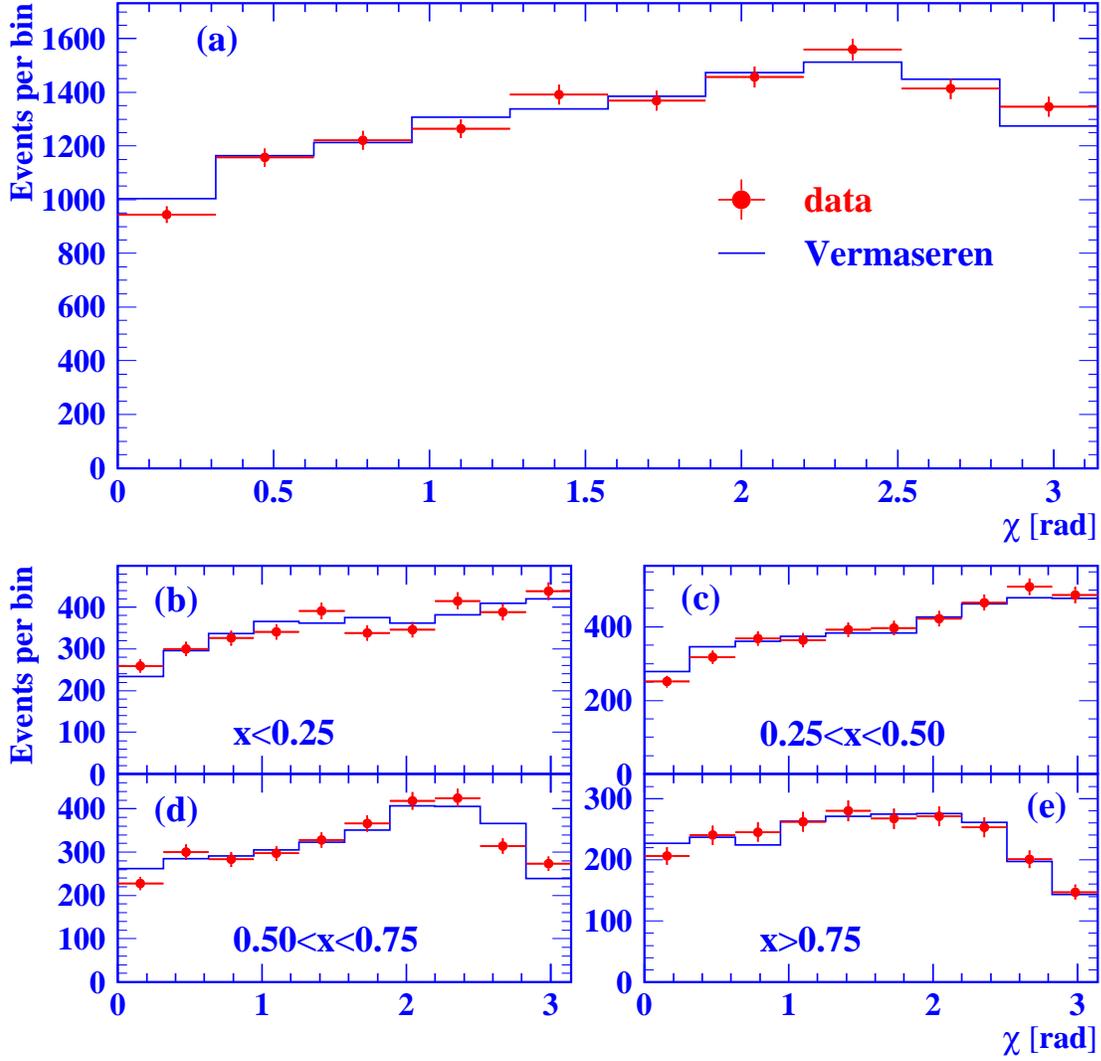,height=15cm}}
\caption{\label{fig:pr271_12}
  Measured azimuthal angle distributions for the combined SW and FD sample
  for different ranges in $x$.
  The points represent the data with the statistical error only, and the full 
  line is the prediction of the Vermaseren Monte Carlo normalised to the 
  number of events in the data
  }
\end{center}
\end{figure}
%
%
\begin{figure}
\begin{center}
\mbox{\epsfig{file=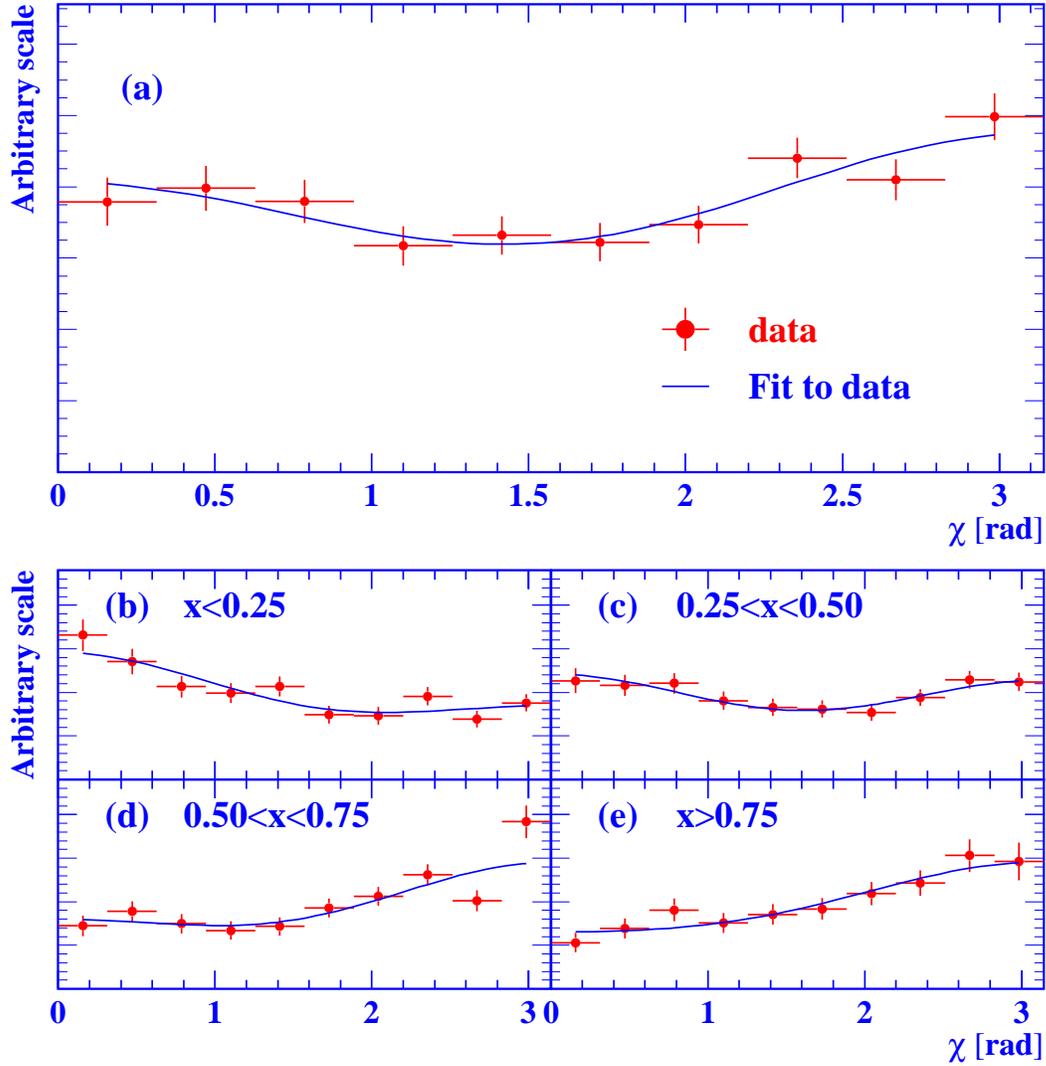,height=15cm}}
\caption{\label{fig:pr271_13}
  Corrected azimuthal angle distributions for the combined SW and FD sample
  for different ranges of $x$.
  The points represent the corrected data with the statistical error only, and 
  the full line is a fit to the data based on Eq.\protect~\ref{eqn:fit}
  }
\end{center}
\end{figure}
%
%
\begin{figure}
\begin{center}
\mbox{\epsfig{file=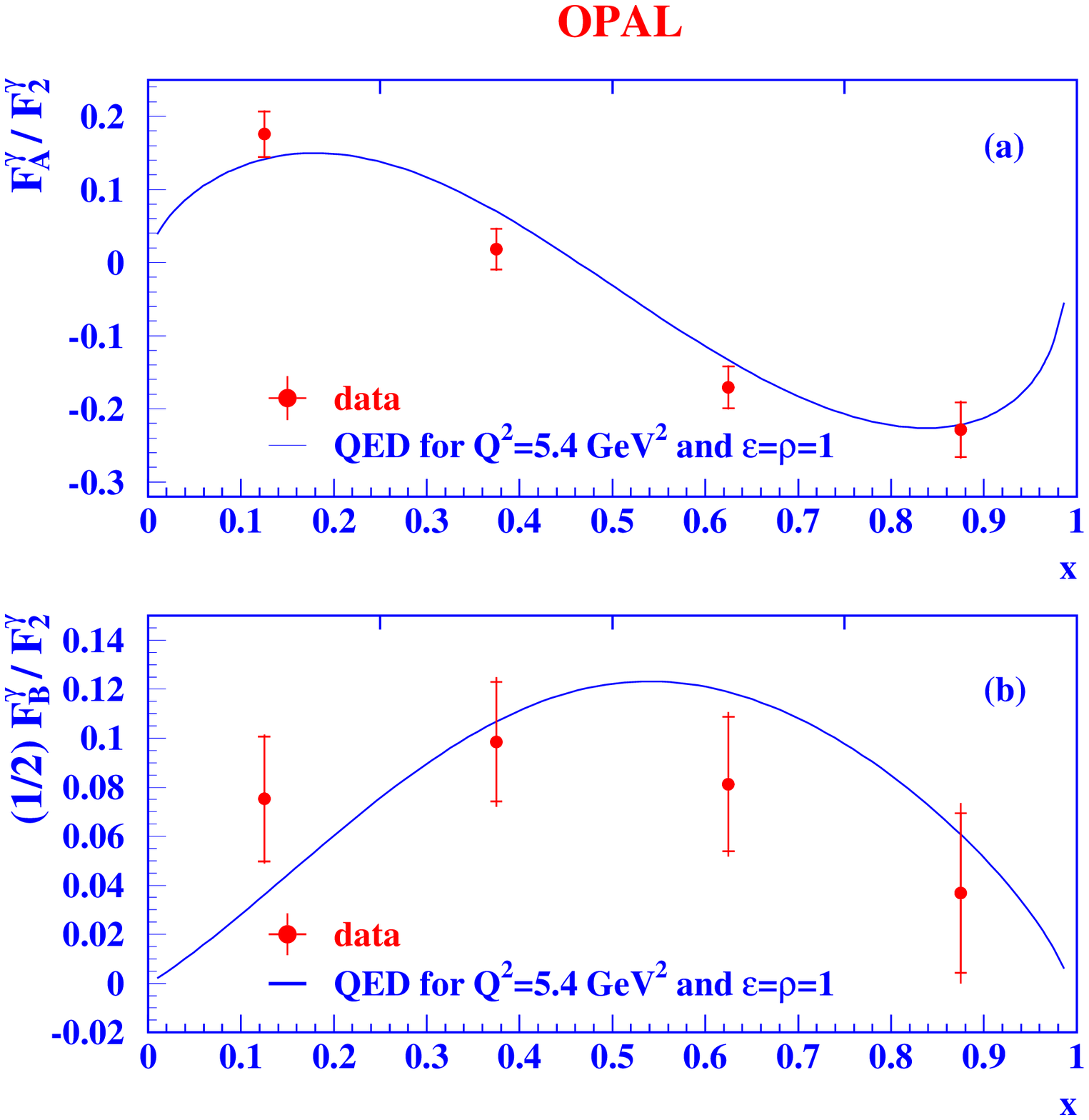,height=15cm}}
\caption{\label{fig:pr271_14}
  The measured structure function ratios \faoft and $\half\fboft$
  for the combined SW and FD sample. 
  The points represent the data with their statistical (inner error bars) 
  and total errors (outer error bars). 
  The solid lines are the QED predictions for 
  $\qsq = 5.4$ \gevsq and $\epsilon=\rho=1$.
  $\chi^2/{\rm dof} = 5.9/4$ for \faoft and $\chi^2/{\rm dof} = 4.4/4$ for 
  $\half\fboft$
  } 
\end{center}
\end{figure}
%
%
\begin{figure}
\begin{center}
\mbox{\epsfig{file=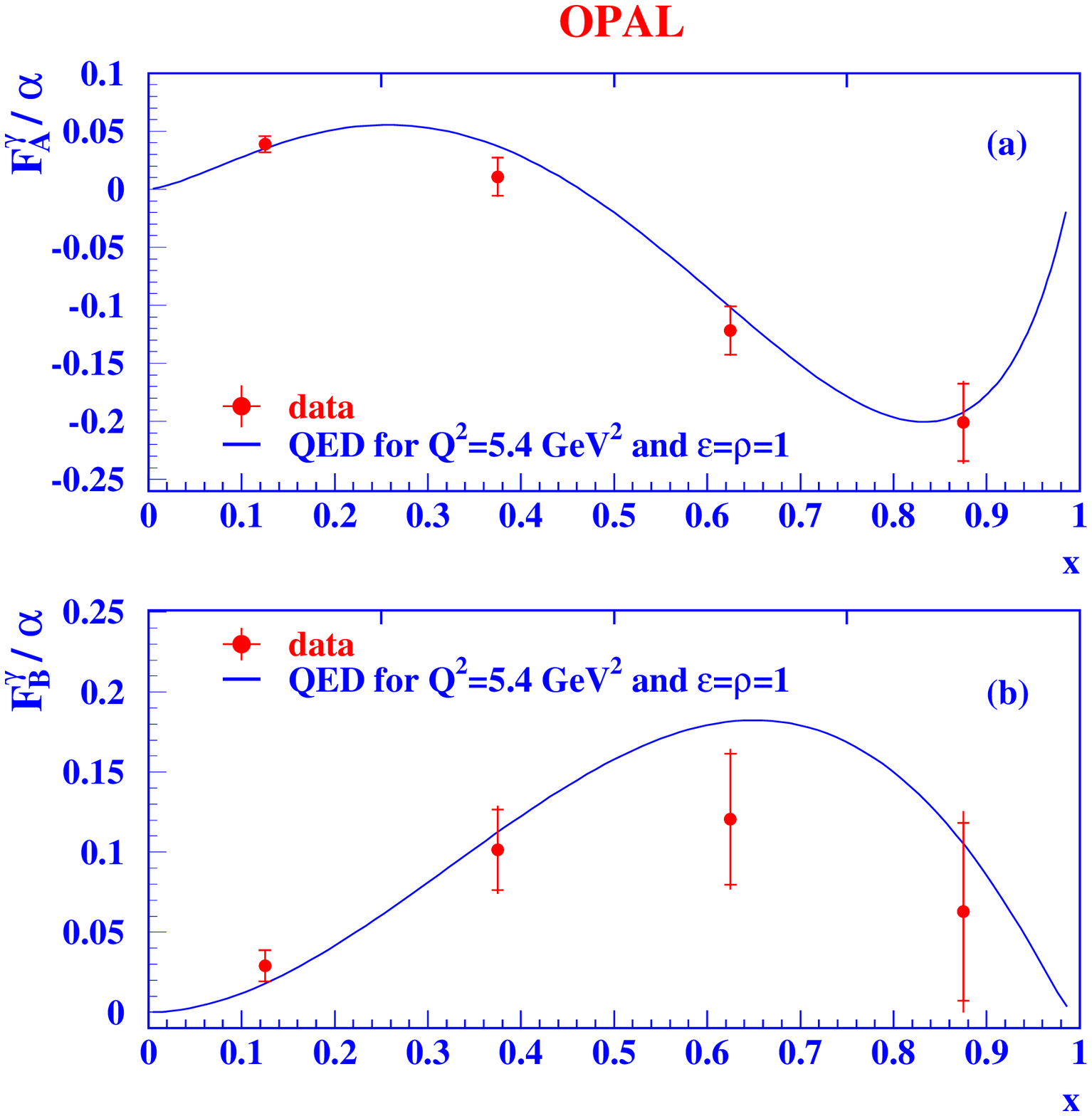,height=15cm}}
\caption{\label{fig:pr271_15}
  The measured structure functions \faqed and \fbqed
  for the combined SW and FD sample. 
  The points represent the data with their statistical (inner error bars) 
  and total errors (outer error bars). 
  The solid lines are the QED  predictions for
  $\qsq = 5.4$ \gevsq and $\epsilon=\rho=1$.
  $\chi^2/{\rm dof} = 3.5/4$ for \faqed and $\chi^2/{\rm dof} = 3.8/4$ for 
  \fbqed
  } 
\end{center}
\end{figure}
%
%

\begin{thebibliography}{10}

\bibitem{BUD-7501}
V.M. Budnev, I.F. Ginzburg, G.V. Meledin, and V.G. Serbo,
\newblock Phys. Rep. {\bf 15}, 181 (1975).

\bibitem{PET-8301}
C.~Peterson, P.M. Zerwas, and T.F. Walsh,
\newblock Nucl. Phys. {\bf B229}, 301 (1983).

\bibitem{CEL-8301TPC-8401PLU-8501}
CELLO Collaboration, 
\newblock H.J. Behrend et~al.,
\newblock Phys. Lett. {\bf 126B}, 384 (1983);\\
TPC/2$\gamma$ Collaboration,
\newblock  M.P. Cain et~al.,
\newblock Phys. Lett. {\bf 147B}, 232 (1984);\\
PLUTO Collaboration,
\newblock C.~Berger et~al.,
\newblock Z. Phys. {\bf C27}, 249 (1985).

\bibitem{OPALPR088}
OPAL Collaboration,
\newblock R.~Akers et~al.,
\newblock Z. Phys. {\bf C60}, 593 (1993).

\bibitem{DEL-9601}
DELPHI Collaboration, 
\newblock P.~Abreu et~al.,
\newblock Z. Phys. {\bf C69}, 223 (1996).

\bibitem{L3C-9801}
L3 Collaboration,
\newblock M.~Acciarri et~al.,
\newblock Phys. Lett. {\bf B438}, 363 (1998).\\

\bibitem{OPALPR182}
OPAL Collaboration,
\newblock K.~Ackerstaff et~al.,
\newblock Z. Phys. {\bf C74}, 49 (1997).

\bibitem{BER-8701}
C.~Berger and W.~Wagner,
\newblock Phys. Rep. {\bf 146}, 1 (1987).

\bibitem{KES-6001}
P.~Kessler,
\newblock Il Nuovo Cimento {\bf 17}, 809 (1960).

\bibitem{WEI-3401WIL-3401}
C.F. von Weizs{\"a}cker,
\newblock Z. Phys. {\bf 88}, 612 (1934);\\
E.J. Williams,
\newblock Phys. Rev. {\bf 45}, 729 (1934).

\bibitem{PLU-8405}
PLUTO Collaboration,
\newblock C.~Berger et~al.,
\newblock Phys. Lett. {\bf 142B}, 119 (1984).

\bibitem{ART-9501}
N.~Arteaga, C.~Carimalo, P.~Kessler, and S.~Ong,
\newblock Phys. Rev. {\bf D52}, 4920 (1995).

\bibitem{SEY-9801}
R.~Nisius, M.H.~Seymour, RAL-TR-1998-079, (1998), hep-ph/9812281, 
to be published in  Phys. Lett. {\bf B}.

\bibitem{AUR-9601}
P.~Aurenche, G.~Schuler, et~al.,
\newblock $\gamma\gamma$ physics,
\newblock in {\em Proceedings of Physics at LEP2 Vol1.}, edited by
  G.~Altarelli, T.~Sj{\"o}strand, and F.~Zwirner, p291, 1996,
\newblock CERN 96-01.

\bibitem{SMI-7701VER-7901VER-8301}
J.~Smith, J.A.M. Vermaseren, and {G. Grammer~Jr.},
\newblock Phys. Rev. {\bf D15}, 3280 (1977);\\
J.A.M. Vermaseren, J.~Smith, and {G. Grammer~Jr.},
\newblock Phys. Rev. {\bf D19}, 137 (1979);\\
J.A.M. Vermaseren,
\newblock Nucl. Phys. {\bf B229}, 347 (1983).

\bibitem{BHA-7701}
R.~Bhattacharya, {G. Grammer~Jr.}, and J.~Smith,
\newblock Phys. Rev. {\bf D15}, 3267 (1977).

\bibitem{BDK-8501BDK-8601BDK-8602BDK-8603}
F.~Berends, P.~Daverveldt, and R.~Kleiss,
\newblock Nucl. Phys. {\bf B253}, 421 (1985);\\
F.~Berends, P.~Daverveldt, and R.~Kleiss,
\newblock Comp. Phys. Comm. {\bf 40}, 271 (1986);\\
F.~Berends, P.~Daverveldt, and R.~Kleiss,
\newblock Comp. Phys. Comm. {\bf 40}, 285 (1986);\\
F.~Berends, P.~Daverveldt, and R.~Kleiss,
\newblock Nucl. Phys. {\bf B264}, 243 (1986).

\bibitem{SCH-9801}
G.A. Schuler,
\newblock Comp. Phys. Comm. {\bf 108}, 279 (1998).

\bibitem{HIL-9301}
J.~Hilgart, R.~Kleiss, and F.L. Diberder,
\newblock Comp. Phys. Comm. {\bf 75}, 191 (1993).

\bibitem{FUJ-9801}
J.~Fujimoto et~al.,
\newblock Comp. Phys. Comm. {\bf 100}, 128 (1997).

\bibitem{OPALPR021ALL-9301ALL-9401AND-9401}
OPAL Collaboration,
\newblock K.~Ahmet et~al.,
\newblock Nucl. Instr. and Meth. {\bf A305}, 275 (1991);\\
P.P. Allport et~al.,
\newblock Nucl. Instr. and Meth. {\bf A324}, 34 (1993);\\
P.P. Allport et~al.,
\newblock Nucl. Instr. and Meth. {\bf A346}, 476 (1994);\\
B.E. Anderson et~al.,
\newblock IEEE Transactions on Nuclear Science {\bf 41}, 845 (1994).

\bibitem{MAR-8801-KNO-8801-CAT-9101-ABB-9001-SEY-9201}
G.~Marchesini and B.R. Webber,
\newblock Nucl. Phys. {\bf B310}, 461 (1988);\\
I.G. Knowles,
\newblock Nucl. Phys. {\bf B310}, 571 (1988);\\
S.~Catani, G.~Marchesini, and B.R. Webber,
\newblock Nucl. Phys. {\bf B349}, 635 (1991);\\
G.~Abbiendi and L.~Stanco,
\newblock Comp. Phys. Comm. {\bf 66}, 16 (1991);\\
M.H. Seymour,
\newblock Z. Phys. {\bf C56}, 161 (1992).

\bibitem{JDH-9401}
S.~Jadach, B.F.L. Ward, and Z.~W\c{a}s,
\newblock Comp. Phys. Comm. {\bf C79}, 503 (1994).

\bibitem{ALL-9201}
J.~Allison et~al.,
\newblock Nucl. Instr. and Meth. {\bf A317}, 47 (1992).

\bibitem{BLO-8401BLO-9601}
V.~Blobel,
\newblock DESY84-118  (1984);\\
V.~Blobel,
\newblock Regularized Unfolding for High-Energy Physics Experiments,
\newblock RUN program manual, unpublished  (1996).

\bibitem{OPALPR185OPALPR207OPALPR213}
OPAL Collaboration,
\newblock K.~Ackerstaff et~al.,
\newblock Z. Phys. {\bf C74}, 33 (1997);\\
OPAL Collaboration,
\newblock K.~Ackerstaff et~al.,
\newblock Phys. Lett. {\bf B411}, 387 (1997);\\
OPAL Collaboration,
\newblock K.~Ackerstaff et~al.,
\newblock Phys. Lett. {\bf B412}, 225 (1997).

\bibitem{BUI-9401}
A.~Buijs, W.G.J. Langeveld, M.H. Lehto, and D.J. Miller,
\newblock Comp. Phys. Comm. {\bf 79}, 523 (1994).

\end{thebibliography}
\end{document}